\documentclass[12pt]{iopart}

\usepackage{graphicx}

\def\Tr{{\rm Tr}}
\begin{document}
\title{Revised Geometric Measure of Entanglement}
\author{Ya Cao, An Min Wang}

\address{Department of Modern Physics, University of Science
and Technology of China, Hefei, 230026, People's Republic of
China}

\ead{anmwang@ustc.edu.cn}
\ead{caoy1209@mail.ustc.edu.cn}

\begin{abstract}
 We present an revised geometric measure of entanglement
(RGME). The revised version is an entanglement monotone. Some
useful inequalities about RGME are deduced. For exemplification,
we give the formulas of RGME for the two-parameter class of states
in $2 \otimes n$ quantum system, the two particles high
dimensional maximally entangled mixed state, the isotropic state
including $n$-particle $d$-level case and two multipartite bound
entangled states. The result shows there is a relation
$\widetilde{E}_{\sin^2} \leq E_{re}$, which indicates that the
RGME is an appropriate measure
of entanglement. %As the RGME's application,

% then make a comparison
%between RGME and other measures.
%The essence of calculating the revised geometric measure of entanglement
%(RGME) is to find the nearest separable state.Then we use the
%revised measure not only to calculate 2-qubit state systems, but
%also to compute the multi-particle cases.
\end{abstract}
\pacs{03.67.Mn, 03.67.-a, 03.65.Ud}



\maketitle

\section{Introduction}
Entanglement, first noted by Einstein-Podolsky-Rosen (EPR) [1] and
Schr\"{o}dinger [2] is an essential feature of quantum mechanics.
To date, the entangled state has become a very useful resource in
many basic problems of quantum computation and quantum
communication. As a result, the task of quantifying entanglement
has emerged as one conspicuous theme in quantum information theory
(QIT). As far as knowledge goes, the quantification of
entanglement is well understood for bipartite pure states, in a
more complex scenario (multipartite systems or mixed states) a
complete theory on the characterization and quantification of
entanglement present even great challenge.

\par In this paper, we present an attempt to explore this challenge
by investigating the amount of entanglement in high dimensional or
multipartite quantum system. The known measures of entanglement
have the entanglement of formation (EOF), the Negativity and the
relative entropy of entanglement(RE) [3-11]. Broadly speaking,
there are two main approaches taken to the definition of
entanglement measures. An operational approach [3], in which the
measures of entanglement are related to physical tasks that one
can perform with a quantum state, as quantum communication, and an
axiomatic approach [8,9,12], which starts from desirable axioms
that a ``good" entanglement measure should satisfy, and then
attempts to construct such measures, for example, the RE [8,9],
the Negativity [10,11] and the robustness of entanglement [12,13]
belong to axiomatic measures. While the entanglement cost [19,20],
the distillable entanglement [21] and the singlet fraction [22,23]
belong to operational measures. One of our authors had ever tried
to suggest a generalization of the EOF [24] and a modification of
the RE [25]. Recently, a multipartite entanglement measure based
on the geometry of Hilbert has been proposed, the geometric
measure of entanglement (GME) [14-18].
%{\em Among these
%measures, the geometric measure of entanglement (GME) [14-18]
%attracts our attention and need further investigation.}

\par As regards the GME, impressive achievements have been obtained,
The merit of this measure is that it is suitable for any-partite
systems with any dimension, although determining it analytically
for generic state remains a challenges. We simply wish to
investigate it further and make it admit a generalisation, that
is, so-called revised GME (RGME). Through the RGME, we quantify
the entanglement of two-parameter class of states in $2 \otimes n$
quantum system, two particles high dimensional maximally entangled
mixed state, isotropic state including $n$-particle $d$-level
case, and two multipartite bound entangled states. Furthermore, we
obtain an important bound relation for these states. Our results
indeed demonstrate the RGME is an appropriate measure of
entanglement.

\par
The paper is organized as follows: in Sec.II, we review the GME.
In Sec.III. we introduce one generalisation called RGME and
investigate its properties in detail. Then we calculate GME, RGME
and other entanglement measures through some special classes of
quantum states in Sec.IV. In Sec.V. we summarize some concluding
remarks.

\section{Geometric measure of entanglement}\label{sec1}
Exploring a geometric approach to quantify measure of entanglement
is first introduced by Shimony [26] in the setting of bipartite
pure states, and  then generalized to the multipartite setting
(via projection operations of various ranks) by Barnum and Linden
[27]. Tzu-ChichWei and Paul M.Goldbart further provide the GME on
the base of their works [14-18]. The GME for pure state
$|\psi\rangle$ is defined as:
\begin{equation}
E_{\sin ^2} = 1 - \Lambda _{\max }^2 = 1 - \mathop {\max
}\limits_{separable\{\phi\} } \left\| {\left\langle {\phi \left|
\psi \right\rangle } \right.} \right\|^2,
\end{equation}

\noindent where $|\phi\rangle$ is a general $n$-partite pure state
with the form (expanded in the local bases ${|e_{p_i}^{\left( i
\right)}\rangle})$
\begin{equation}
\left| \phi \right\rangle = \otimes_{i = 1}^n \left| {\phi ^{\left(
i \right)}} \right\rangle = \sum\limits_{p_1 p_2 \cdots p_n }
%{C_{p_1 }^{\left( 1 \right)} C_{p_2 }^{\left( 2 \right)} \cdots
%C_{p_n}^{\left( n \right)}
\chi_{p_1p_2 \cdots p_n}\left| {e_{p_1 }^{\left( 1 \right)} e_{p_2
}^{\left( 2 \right)} \cdots e_{p_n}^{\left( n \right)} }
\right\rangle
\end{equation}
In basis independent form, we have
\begin{equation}
\langle\psi (\otimes_{j(\neq
i)}^{n})|\phi^{(j)}\rangle)=\Lambda\langle\phi^{(i)}|,
~~~(\otimes_{j (\neq i)}^{n}\langle
\phi^{(j)})|\psi\rangle=\Lambda|\phi^{(i)}\rangle.
\end{equation}
which are independent of the choice of the local basis. The
physical meaning of the GME can be seen from entanglement
eigenvalue $\Lambda _{\max }$ which is the cosine of angle between
the pure state and its closest separable state. Of course, the
stronger the entanglement of state becomes, the farer its closest
separable state will be, and the larger will be the angle between
them. We remark that determining the entanglement of
$|\psi\rangle$ is equivalent to finding the Hartree approximation
to the ground state of the auxiliary Hamiltonian $H=-|\psi\rangle
\langle \psi|$ [15]. The extension to mixed state can be made via
the use of the convex roof (or hull) construction as done for EOF.
The essence of problem is a minimization over all decompositions
into pure states, i.e.
\begin{eqnarray}
 & &\rho = \sum\limits_i {p_i } \left| {\psi _i } \right\rangle \left\langle
{\psi _i } \right|,\nonumber \\
 & &E\left( \rho \right) = \left( {coE_{pure} } \right)\left( \rho \right) =
\mathop {\min }\limits_{\left\{ {p_i ,\psi _i } \right\}}
\sum\limits_i {p_i } E_{pure} \left( {\left| {\psi _i }
\right\rangle } \right).
\end{eqnarray}

\noindent so, for the general mixed state, it is difficult to write
out the clear analytical expression of the GME. It is worth
indicating that arbitrary two-qubit mixed state, its GME has been
given [19]:
\begin{equation}
E_{\sin ^2} = \frac{1}{2}\left( {1 - \sqrt {1 - C\left( \rho
\right)^2} } \right).
\end{equation}
where $C(\rho)$ is concurrence of an arbitrary two-qubit mixed
state defined as follows:
\begin{equation}
C(\rho)=\max {  \{0,
\sqrt{\lambda_1}-\sqrt{\lambda_2}-\sqrt{\lambda_3}-\sqrt{\lambda_4}\}
}
\end{equation}
and $\lambda_1 \geq \lambda_2 \geq \lambda_3\geq \lambda_4$ are
the square roots of eigenvalues of the product
$\rho_{AB}\tilde{\rho}_{AB}$,
\begin{equation}
\tilde{\rho}_{AB}=(\sigma_{y}\otimes
\sigma_{y})\rho_{AB}^{\ast}(\sigma_{y}\otimes \sigma_{y})
\end{equation}

 For the sake of clarifying, we use the GME to calculate some
simple 2-qubit states.

\noindent Example 1 : \begin{equation} \rho _1 = \lambda \left|
{\Phi ^\dag } \right\rangle \left\langle {\Phi ^\dag } \right| +
\left( {1 - \lambda } \right)\left| {01} \right\rangle \left\langle
{01} \right|
\end{equation}
where $| {\Phi ^\dag } \rangle=
\frac{1}{\sqrt{2}}(|00\rangle+|11\rangle$).

\begin{equation}
E_{\sin ^2} = \frac{1}{2}\left( {1 - \sqrt {1 - C\left( \rho
\right)^2} } \right) = \frac{1}{2}\left( {1 - \sqrt {1 - |\lambda|
^2} } \right).
\end{equation}

\noindent Example 2 :
\begin{equation}
\rho_2 = A\left| {01} \right\rangle \left\langle {01} \right| +
\left( {1 - A} \right)\left| {10} \right\rangle \left\langle {10}
\right| + \displaystyle\frac{G}{2}\left( {\left| {01} \right\rangle
\left\langle {10} \right| + \left| {10} \right\rangle \left\langle
{01} \right|} \right)
\end{equation}

\noindent where $G$ satisfies $G \le 2\sqrt {A\left( {1 - A}
\right)} $. This condition ensures the state $\rho_2$ is
semi-definite. Negativity and Concurrence of this class of states
are equal, {\it i.e.} $ N\left( \rho_2 \right) = C\left( \rho_2
\right) = G $. It's easy to calculate the GME,
\begin{equation}
E_{\sin ^2} = \frac{1 - \sqrt {1 - G^2} }{2} .
\end{equation}

%%%%%From the definition of GME, for pure state, its closest separable
%%%%%state is also pure state $|\phi\rangle$, {\it i.e.} Eq.(2). However,
%%%%%generally speaking, even in the case of pure states, most of its
%%%%%closest separable states are mixed states, this standpoint has been
%%%%%presented in many Refs.[8,9]. So the definition of GME exists flaws.
%%%%%Only some special states, their closest separable states can be pure
%%%%%states, such as examples in Refs.[15,17,18]. In addition, as far as
%%%%%a mixed state is concerned, the convex structure is complicated to
%%%%%compute, so the above definition Eq.(3) is inapposite in this sense.
%%%%%Due to these flaws, one main purpose of this paper is to revise GME
%%%%%to overcome them.

Based on the requirement of calculating GME of pure state, its
closest separable pure state $|\phi\rangle$ is given, {\it i.e.}
Eq.(2). However, generally speaking, even in the case of pure
states, most of its closest separable states are mixed states,
this standpoint has been presented in many Refs.[8,9]. Only some
special states, their closest separable states can be pure states,
such as examples in Refs.[15,17,18]. Thus it is necessary to
generalize the original definition to reach the perfectness. In
addition, as far as a mixed state is concerned, the convex
structure is complicated to compute because it adds the amount of
calculation and the level of difficulty. Above definition Eq.(3)
is a common method to deal with generalisation of mixed state
which obviously poses a challenge to compute from the sense of
complication of computation. Whether there is a different method
which can solve this problem to make it compute easily and fills
with physical meaning. This is one aim of this paper. Concretely,
the ''flaws" of the GME is the limitation of the expression of the
closest separable state. It's far from enough to just consider
Eq.(2) because the closest separable state varies with different
initial state generally. Due to this incompletement, our main
purpose of this paper is to generalize GME to avoid facing this
embarrassment.

\section{Revised geometric measure of entanglement}\label{sec2}
The motivation for constructing the GME is to address the degree
of entanglement from a geometric viewpoint, regardless of the
number of parties. Yet, there is some room to generalize in the
original definition of GME. In this section, we propose the
revised GME (RGME) which just a generalisation to make it perfect
and we elucidate the revision by some concrete examples.

\subsection{Definition}
We begin with the revision of GME, which is defined as£º
\begin{equation}
\widetilde{E}_{\sin ^2} ( \rho ) = \min\limits_{\sigma\in S} ( {1 -
F^2( {\rho ,\sigma } )} ),
\end{equation}

\noindent where
\begin{equation}
\;F\left( {\rho ,\sigma } \right)={\rm tr}\sqrt {\rho
^{\frac{1}{2}}\sigma \rho ^{\frac{1}{2}}} .
\end{equation}

\noindent $S$ denotes the set of separable states. Comparing
Eq.(12) with Eq.(1), we obtain the relation,
\begin{equation}
\Lambda _{\max } ^2 = \mathop {\max }\limits_{\sigma \in S}
F^2\left( {\rho ,\sigma } \right).
\end{equation}
Remark when density matrices $\rho ,\sigma $ represent pure
states, Fidelity equals to overlap, above formula can reduce to
the definition of the GME {\it i.e.} Eq.(1). The maximum of
overlap is totally different from the maximal Fidelity in the
sense the latter's variational range is wider than the form's
case, thus our revised GME is more appropriate. The essence of
proposed RGME is calculating the Fidelity between given state and
its closest separable state. Finally, it reduces to the search of
the closest separable state.

Because the relation between Bures metric and Fidelity, and the
fact that Bures metric is positive, we have
\begin{equation}
\quad Bures\left( \rho \right) = \sqrt {1 - F^2\left( {\rho ,\sigma
} \right)}   .
\end{equation}

\noindent then the RGME can be expressed as £º
\begin{equation}
\widetilde {E}_{\sin ^2} \left( \rho \right) = \mathop {\min
}\limits_{\sigma
\in S} \left( { Bures^2(\rho)} \right). \\
\end{equation}

Let us see whether the RGME $\widetilde {E}_{\sin ^2} $ is a good
entanglement measure or not? We know a good entanglement measure
should satisfy some properties [28,29]. It's easy to prove the
RGME indeed satisfies these requirements.
%%%%%%%%% proof of the good entanglement monotone %%%%%%%%%%%%%%%%%%%%%%%%%%%%%%

Now, we verify it non-increasing under local operation and
classical communication (LOCC) transformation using the Uhlmann'
theorem [30]. Proof: Assume $\varepsilon $ is a trace-preserving
quantum operation, $\rho ,\sigma $ are density operators. Let
$\left| \psi \right\rangle ,\left| \varphi \right\rangle $ be
purifications of $\rho ,\sigma $ in a joint system RQ such that
$F\left({\rho ,\sigma} \right)=\left\langle \psi \right|\left.
\varphi \right\rangle $. Introduce a model environment $E$ for the
quantum operation $\varepsilon $ which starts in a pure state
$\left| 0 \right\rangle $, and interacts with the quantum system Q
via a unitary interaction U. Note $U\left| \psi \right\rangle
\left| 0 \right\rangle $ is a purification of $\varepsilon \left(
\rho \right)$, and $U\left| \varphi \right\rangle \left| 0
\right\rangle $ is a purification of $\varepsilon \left( \sigma
\right)$. By Uhlmann's theorem, we have

\begin{equation}
\begin{array}{l}
 F\left( {\varepsilon \left( \rho \right),\varepsilon \left( \sigma \right)}
\right) \ge \left| {\left\langle \psi \right|\left\langle 0
\right|UU^\dag \left| \varphi \right\rangle \left| 0 \right\rangle
} \right| = \left| {\left\langle \psi \right|\left. \varphi
\right\rangle } \right| = F\left(
{\rho ,\sigma } \right) \\
% - F\left( {\varepsilon \left( \rho \right),\varepsilon \left( \sigma
%\right)} \right) \le - F\left( {\rho ,\sigma } \right) \\
 1 - F\left( {\varepsilon \left( \rho \right),\varepsilon \left( \sigma
\right)} \right) \le 1 - F\left( {\rho ,\sigma } \right) \\
 \tilde {E}_{\sin ^2} \left( {\varepsilon \left( \rho \right)} \right) \le
\tilde {E}_{\sin ^2} \left( \rho \right) \\
 \end{array}
\end{equation}

\noindent Thus the proof finishes.

%%%%%%%%%%%%%%%%%%%%%%%%% LU invariant %%%%%%%%%%%%%%%%%%%%%%%
As for LU invariant, it is determined by the property of Fidelity.
Fidelity is invariant under Local unitary (LU) transformation.
\begin{equation}
\begin{array}{l}
 F\left( {U\rho U^\dag ,U\sigma U^\dag } \right) = tr\sqrt {\left( {U\rho
U^\dag } \right)^{\frac{1}{2}}U\sigma U^\dag \left( {U\rho U^\dag
}
\right)^{\frac{1}{2}}} \\
 \quad \quad \quad \quad \quad \quad \quad
% = tr\left( {\sqrt {U\rho U^\dag }
%} \right)^{\frac{1}{2}}\sqrt {U\sigma U^\dag } \left( {\sqrt {U\rho U^\dag }
%} \right)^{\frac{1}{2}} \\
 \end{array}
\end{equation}

\begin{equation}
\begin{array}{l}
tr\sqrt {\left( {U\rho U^\dag } \right)^{\frac{1}{2}}U\sigma
U^\dag \left( {U\rho U^\dag }
\right)^{\frac{1}{2}}} \\
\quad \quad = tr \sqrt {U\sqrt\rho U^\dag U \sigma U^\dag U\sqrt \rho U^\dag }\\
\quad \quad = tr \sqrt {U\sqrt\rho  \sigma \sqrt \rho U^\dag }\\
\quad \quad = tr U\sqrt {\sqrt\rho  \sigma \sqrt \rho}U^\dag\\
\quad \quad = tr \sqrt {\sqrt\rho  \sigma \sqrt \rho}\\
\quad \quad = F\left( {\rho ,\sigma } \right) \\
\end{array}
\end{equation}
Therefore, we show the RGME is a good entanglement measure.
%%%%%%%%%%%%%%%%%%%%%%%%%%%%%%%%%%%%%%%%%%%%%%%%%%%%%%%%%%%%%%%%%%%%%%%%

One of virtues of the geometric approach to entanglement is its
straightforward adaptability to arbitrary multipartite state (of
finite dimensions). The revision of the GME has similar character.
There are four differences deserved emphasizing between the RGME
and the GME: 1. The RGME use the Fidelity to substitute the
overlap, then whatever the given state is pure or mixed state, in
light of the relation between Fidelity and overlap, the RGME
always can be expressed in Fidelity form congruously. 2. The
revised form abandons the condition that the closest separable
state has the form Eq.(2), even for the case of pure state, say
nothing of the mixed state scenario. 3. The revised version do not
need the convex hull to consider the case of mixed state like GME
which complicates the task of determining mixed-state
entanglement, whereas the essence of problem is attributed to find
out the closest separable state. 4. For the case of pure state
$\rho$, there always exists a bound condition $\widetilde
{E}_{\sin ^2} \left( \rho \right)\leq E_{\sin ^2} \left( \rho
\right) $.

\subsection{Examples}
In this subsection, we use the RGME to re-calculate the foregoing
two examples, the figures are shown for the convenience of analysis.
Therein, the closest separable state of these examples are given and
testified by the method given in Ref.[9].

\noindent Example 1, its closest separable state reads as
following£º
\begin{eqnarray}
\sigma _1 &=& \frac{\lambda }{2}\left( {1 - \frac{\lambda }{2}}
\right)\left| {00} \right\rangle \left\langle {00} \right| +
\frac{\lambda }{2}\left( {1 - \frac{\lambda }{2}} \right)\left(
{\left| {00} \right\rangle \left\langle {11} \right| + \left| {11}
\right\rangle \left\langle {00} \right|} \right)\nonumber
\\
&&+ \left( {1 - \frac{\lambda }{2}} \right)^2\left| {01}
\right\rangle \left\langle {01} \right| + \frac{\lambda ^2}{4}\left|
{10} \right\rangle \left\langle {10} \right| + \frac{\lambda
}{2}\left( {1 - \frac{\lambda }{2}} \right)\left| {11} \right\rangle
\left\langle {11} \right|
\end{eqnarray}

\noindent the RGME is£º
\begin{equation}
\widetilde {E}_{\sin ^2} = 1 - F_{\max }^2 = 1 - \left[ {\left( {1 -
\frac{\lambda }{2}} \right)\sqrt {1 - \lambda } + \lambda \sqrt {1 -
\frac{\lambda }{2}} } \right]^2 .
\end{equation}

Ref.[9] gives its analytical expression of RE
\begin{equation}
E_{re} (\rho_{1})=(\lambda-2)\log{(1-\frac{\lambda} {2} )}
+(1-\lambda)\log(1-\lambda) .
\end{equation}

\noindent Note that in the whole paper we reckon that the formula of
the RE is $E_{re}= \min\limits_{\sigma\in S}{\rm
tr}(\rho\log\rho-\rho\log\sigma)$, where $\log$ denotes logarithm
whose base is two. Now, we analyze the relation about the GME, the
RGME and the RE by virtue of fig.1. From the figure, we see the
value of the RE is larger than other two entanglement measures. At
the same time, the curve of the GME almost superposes to that of the
RGME, which to some extent illustrates the RGME is reasonable with
regard to the GME.

\begin{figure}%[htbp]
\centering
\includegraphics[scale=0.87]{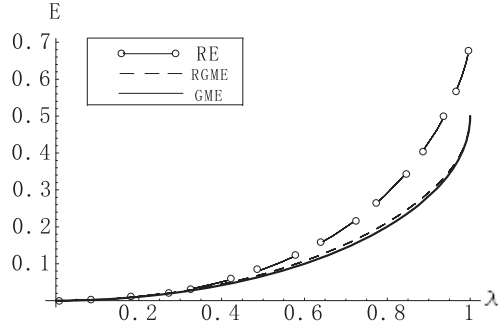} \caption{ The curves of
the GME and RGME almost coincide and lie below the curve of the RE
which show the RGME is better to measure the amount of
entanglement of this state.}
\end{figure}

\noindent Example 2, its closest separable state is
\begin{equation}\sigma_2 = A\left| {01} \right\rangle \left\langle {01}
\right| + \left( {1 - A} \right)\left| {10} \right\rangle
\left\langle {10} \right|\end{equation}

\noindent the RGME is £º\begin{equation}  \fl \widetilde {E}_{\sin
^2} = 1 - \left( {\begin{array}{l}
 \sqrt {\displaystyle \frac{1}{2}\left( {1 - 2A + 2A^2 - \sqrt {1 - 4A + 4A^2 + AG^2 -
A^2G^2} } \right)} \\
 + \sqrt {\displaystyle \frac{1}{2}\left( {1 - 2A + 2A^2 + \sqrt {1 - 4A + 4A^2 + AG^2 -
A^2G^2} } \right)} \\
 \end{array}} \right)^2  .
\end{equation}

\noindent when choosing $A$ as different random numbers, we get
different curves of the RGME showed in fig.2. Obviously, the RGME
tends to GME with the increase of $A$.  when $A\rightarrow1$, the
RGME superposes to the GME .

\begin{figure}%[htbp]
\centering
\includegraphics[scale=0.46]{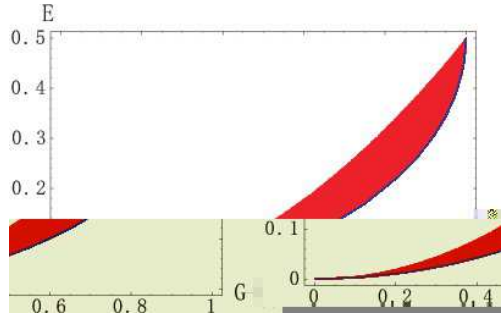}\caption{The red field
denotes different RGME curves for different random number A; the
black line represents the GME. The GGME and the GME are coincident
when $A\rightarrow1$.}
\end{figure}

In above all examples, the closest separable state of the RGME is
identical to that of the RE, and they are mixed states. Yet, we
must emphasize it is not the case in the general situation.
Without loss of generality, we are concerned with the state
\begin{eqnarray}
\rho &=& \alpha ^2\left| {00} \right\rangle \left\langle {00}
\right| + \alpha \sqrt {1 - \alpha ^2} \left( {\left| {00}
\right\rangle \left\langle {11} \right| + \left| {11} \right\rangle
\left\langle {00} \right|} \right) + \left( {1 - \alpha ^2}
\right)\left| {11} \right\rangle \left\langle {11}
\right| \nonumber\\
 & =& \left( {\alpha \left| {00} \right\rangle + \sqrt {1 - \alpha ^2}
\left| {11} \right\rangle } \right)\left( {\alpha \left\langle {00}
\right| + \sqrt {1 - \alpha ^2} \left\langle {11} \right|}
\right)\nonumber\\& =& \left| \xi \right\rangle \left\langle \xi
\right|,
\end{eqnarray}

\noindent where $\alpha \in \left[ {0,1} \right]$. Its closest
separable state under the RE is
\begin{equation}
\sigma = \alpha ^2\left| {00} \right\rangle \left\langle {00}
\right| + \left( {1 - \alpha ^2} \right)\left| {11} \right\rangle
\left\langle {11} \right| ,
\end{equation}

\noindent Naturally, the RE for this state is
\begin{equation}
E_{re} = - \alpha ^2\log \alpha ^2 - (1 - \alpha ^2)\log (1 - \alpha
^2) .
\end{equation}

\noindent However, the closest separable state under the RGME is
\begin{equation}
\fl \sigma^\prime =
\left(1-\sqrt{\frac{1+\sqrt{1-4\alpha^2(1-\alpha^2)}}{2}}\right)\left|
{00} \right\rangle \left\langle {00} \right| +
\sqrt{\frac{1+\sqrt{1-4\alpha^2(1-\alpha^2)}}{2}}\left| {11}
\right\rangle \left\langle {11} \right|,
\end{equation}
By calculation, we know $\sigma^\prime$ is a disentangled state
without reference to $\alpha$, because all eigenvalues of
${\sigma^\prime}^ {T_B}$ are non-negative (PPT criterion).
Accordingly, the RGME is
\begin{equation}
\widetilde {E}_{\sin ^2} = 1 - F_{\max }^2 = \frac{1}{2}\left( {1 -
\sqrt {1 - 4\alpha ^2\left( {1 - \alpha ^2} \right)} } \right),
\end{equation}

\noindent which is equal to the GME
\begin{equation}
E_{\sin ^2} = \frac{1}{2}\left( {1 - \sqrt {1 - 4\alpha ^2\left( {1
- \alpha ^2} \right)} } \right)
%= \frac{1}{2}\left( {1 - \sqrt {1 - C\left( \rho
%\right)^2} } \right)
\\
 = \left\{
\begin{array}{*{20}c}
 1 - \alpha ^2 \ \ & \ \ \left( {\displaystyle\frac{\sqrt 2 }{2} < \alpha < 1 }
\right). \\
 \alpha ^2 \ \ & \ \ \left( {0 \le \alpha \le \displaystyle\frac{\sqrt 2 }{2}} \right).\\
\end{array}  \right .
\end{equation}

\noindent If we use the $\sigma^\prime$ to re-calculate the RE, we
get the relation $E_{re}(\rho, \sigma) \le E_{re}(\rho
,\sigma^\prime)$ which indicates the closet separable state is
indeed the state $\sigma$ under the RE. In order to demonstrate
their relations explicitly, we show the fig.3.

\begin{figure}%[htbp]
\centering
\includegraphics[scale=0.87]{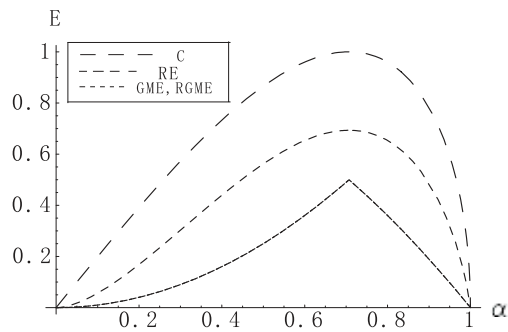} \caption{ The figures
about the RE, the Concurrence, the GME and the RGME are showed, in
which the curve representing RGME superposes to the curve of GME.
Obviously, there is a relation $\widetilde {E}_{\sin ^2}={E}_{\sin
^2} \leq E_{re} \leq Concurrence.$}
\end{figure}

\subsection{Properties of RGME}
It is important to investigate the properties of RGME deeply. In
this subsection, we give some propositions with regard to the
inequality relations about some measures of entanglement.

\noindent \textbf {Proposition 1}. The RGME and the Fidelity satisfy
an universal relation for any state $\rho, \sigma$
\begin{equation}
1 - F\left( {\rho ,\sigma } \right) \le \sqrt{\widetilde {E}_{\sin
^2} \left( \rho \right)}.
\end{equation}

\noindent Proof. Make use of the mathematical formula $\{1-x \leq
\sqrt{1-x^2} , ~~0 \leq x\leq 1\}$. One can see quickly for any
state $\rho,\sigma$
\begin{equation}
 {1 - F\left( {\rho ,\sigma } \right)}  \le \sqrt {1
- F^2\left( {\rho ,\sigma } \right)} .
\end{equation}

\noindent from which and the definition of the RGME, Eq.(9), the
proposition follows.

\noindent \textbf{Proposition 2.} The RGME for bipartite pure states
is smaller than the entanglement of formation or the relative
entropy of entanglement, {\it i.e.}
\begin{equation}
\widetilde{E}_{\sin ^2}(\rho) \leq E_f(\rho)=E_{re}(\rho)
\end{equation}
and the equality is valid only when $\rho$ is separable state.

\noindent Proof. when $\rho=| \psi \rangle \langle \psi|$, and
$\left| \psi \right\rangle$ is decomposed as $A$ and $B$ parts, we
have
\begin{equation}
Bures\left( \rho \right) = \sqrt {1 - F^2\left( {\rho ,\sigma }
\right)} \le - \frac{1}{2}{\rm tr}\left( {\rho _A \log \rho _A }
\right) = \frac{1}{2}S_A\left( \rho \right) ,
\end{equation}

\noindent where $S_A(\rho)$ is the von Neumann reduced entropy. In
term of Eq.(26), then we have
\begin{equation}
1 - F\left( {\rho ,\sigma } \right) \le \frac{1}{2}S_A\left( \rho
\right)
\end{equation}
under the condition that $\rho $ is bipartite pure state,
accordingly, we can deduce that the RGME satisfies the inequality
\begin{equation}
\widetilde{E}_{\sin ^2}{(\rho)} \leq 1-
\left(1-\frac{S_A{(\rho)}}{2}\right)^2 =
S_A{(\rho)}-\frac{S^2_A{(\rho)}}{4}\leq S_A{(\rho)}
\end{equation}
and the equality is valid only $S_A(\rho)$ is zero, that is, $\rho$
is separable.

Due to the existence of Schmidt decomposition [31] in bipartite pure
state system, the EF is equal to the von Neumann reduced entropy and
the RE, {\it i.e} $S_A{(\rho)}=E_f{(\rho)}=E_{re}{(\rho)}$, thus, we
arrive at the desired relation.

\noindent \textbf{Proposition 3.} The RGME is smaller than the trace
distance for any pure state $| \psi \rangle$, {\it i.e.}
\begin{equation}
\widetilde{E}_{\sin ^2} \left( {\left| \psi \right\rangle } \right)
\le D( {|\psi \rangle ,\sigma }) .
\end{equation}

\noindent Proof. In the case of pure state $\rho=| \psi \rangle
\langle \psi|$, there exits another relation
\begin{equation}
1 - F^2\left( {\left| \psi \right\rangle ,\sigma } \right) \leq
\frac{1}{2}{\rm tr}\left( {\left| {\left| \psi \right\rangle
\left\langle \psi \right| - \sigma } \right|} \right)=D( {|\psi
\rangle ,\sigma }),
\end{equation}

\noindent where $D(|\psi \rangle ,\sigma )$ is trace distance
between state $|\psi \rangle$ and $\sigma$. In light of the
definition of the RGME, we obtain the sought inequality finally.

As we know, among many measures of entanglement, the von Neumann
entropy is very important and it has extensive application. Eq.(30)
and (31) prompt us to ask what's the relation between $D(
{\rho,\sigma })$ and $S\left( \rho \right)$?

\noindent \textbf{Proposition 4.} The von Neumann entropy and
distance trace satisfy the inequality
\begin{equation}
{S\left( \rho \right)} \leq 2D(\rho, \sigma)+\frac{1}{e} .
\end{equation}
under the condition that $\sigma$ is pure state and the dimension of
any state $\rho$ fulfills $d \geq 4$.

\noindent Proof. Due to \textbf{Fannes' inequality}: suppose $\rho $
and $\sigma $ are density matrices such that the trace distance
between them satisfies $\frac{1}{2}{\rm tr}\left( {\left| {\rho -
\sigma } \right|} \right) \le \frac{1}{e}$, then
\begin{equation}
\left| {S\left( \rho \right) - S\left( \sigma \right)} \right| \le
\frac{1}{2}{\rm tr}\left| {\rho - \sigma } \right| \cdot \log d +
\eta \left( {\frac{1}{2}{\rm tr}\left| {\rho - \sigma } \right|}
\right) ,
\end{equation}

\noindent where $d$ is the dimension of the Hilbert space, and $\eta
\left( x \right) = - x\log x$. Removing the restriction that
$\frac{1}{2}{\rm tr}\left( {\left| {\rho - \sigma } \right|} \right)
\le \frac{1}{e}$, there is an inequality
\begin{equation}
\left| {S\left( \rho \right) - S\left( \sigma \right)} \right| \le
\frac{1}{2}{\rm tr}\left| {\rho - \sigma } \right| \cdot \log d +
\frac{1}{e}.
\end{equation}

\noindent Using the weaker inequality, when $ S ( \sigma ) = 0$,
{\it i.e.} $\sigma $ is a pure state, it yields
\begin{equation}
\left| {S\left( \rho \right)} \right| \le \frac{1}{2}{\rm tr}\left|
{\rho - \sigma } \right| \cdot \log d + \frac{1}{e} ,
\end{equation}

%\noindent omitting $\frac{1}{e}$ and using the fact that the von
%Neumann entropy is non-negative, so
\noindent Note that in the definition - and throughout this paper-
logarithms indicated by ``log" are taken to base two, while `ln'
indicates a natural logarithm. When $\frac{\log d}{2} \geq 1$, we
deduce the relation $d \geq 4$, then we obtain a relation between
the von Neumann entropy and distance trace
\begin{equation}
{S\left( \rho \right)}  \le {\rm tr}\left| {\rho - \sigma }
\right|=2D(\rho, \sigma)+\frac{1}{e} .
\end{equation}
hence the proposition is proved.

As a subsidiary product, we combine above deduction with the facts
\begin{equation}
F(\rho,\sigma) + D(\rho,\sigma) \geq 1,~~~  F^2(\rho,\sigma) +
D^2(\rho,\sigma) \le 1 ,
\end{equation}
then the unambiguous relation under the condition $\rho$ is pure
state becomes
\begin{equation}
(1-F)^2 \le 1-F \le \widetilde{E}_{\sin ^2} \le 1-F^2 \le D \le
\sqrt{1-F^2} \le \sqrt{D} .
\end{equation}

Thus, we have investigated the relations between different measures
of entanglement. In view of their different physical meaning for
measuring the amount of entanglement, we believe these relations may
imply much in many problems such as comparison about different
measures of entanglement, discussion about the bound of different
measures of entanglement.

\section{RGME of some special classes of states}\label{sec3}
Progress in the quantification of entanglement for a mixed state has
resided primarily in the domain of bipartite systems. If we can
formulate the universal measures of many-particle system and
multi-partite system entanglement, they would have many applications
[32]. One purpose of this paper is to achieve some analytical form
of the RGME for some special cases that are interesting in theory.
Concretely, we use mathematical induction method to obtain the
expressions of the RGME for two-parameter class of states in $2
\otimes n$ quantum system, bipartite maximally entangled mixed
state, isotropic state including $n$-particle $d$-level case, and
two multipartite bound entangled states, the relation between the
RGME and the corresponding GME are also obtained. At the same time,
we obtain an important conclusion that the RE is an upper bound on
the RGME for these states. Based on these results, we can see the
advantages of RGME relative to other measures of entanglement
clearly.

\subsection{RGME of two-parameter class of states in $2 \otimes n$ quantum system}

Now, we consider the class of states with two real parameters
$\alpha$ and $\gamma$ in $2 \otimes n$ quantum system. A finite
dimensional truncation of a single two level atom interacting with a
single-mode quantized field [33] can be regarded as a $2 \otimes n$
quantum system.

Firstly, we deal with the simple $n=3$ case. For $2 \otimes 3$
quantum system, two parameters state [34] can be expressed as:
\begin{eqnarray}
 \tilde {\rho } &=& \beta \left| {\psi ^ + } \right\rangle \left\langle
{\psi ^ + } \right| + \gamma \left| {\psi ^ - } \right\rangle
\left\langle {\psi ^ - } \right| + \beta \left( {\left| {00}
\right\rangle \left\langle {00} \right| + \left| {11} \right\rangle
\left\langle {11} \right|} \right)\nonumber
\\
&& + \alpha \left( {\left| {02} \right\rangle \left\langle {02}
\right| + \left| {12} \right\rangle \left\langle {12} \right|}
\right),
\end{eqnarray}

\noindent where \begin{equation} |\phi^{\pm}\rangle=
\frac{1}{\sqrt{2}}(|00\rangle \pm |11\rangle),
~~~~|\psi^{\pm}\rangle= \frac{1}{\sqrt{2}}(|01\rangle \pm
|10\rangle).
\end{equation}
\noindent and $\beta$ is dependent on $\alpha$ and $\gamma$ by the
unit trace condition,
\begin{equation}
 2\alpha + 3\beta + \gamma = 1,~~~ 0 \le \alpha \le \frac{1}{2} .
\end{equation}

\noindent remark when $ \frac{1}{2} \le \gamma \le 1$, the state is
entangled. Similar to the method in Ref.[35], we know the closest
separable state of $\tilde {\rho }$ has the following form£º
\begin{eqnarray}
 \tilde {\sigma }^ * &=& p_{\,1} \left| {\psi ^ + } \right\rangle \left\langle
{\psi ^ + } \right| + p_{\,2} \left| {\psi ^ - } \right\rangle
\left\langle {\psi ^ - } \right| + p_{\,3} \left| {00} \right\rangle
\left\langle {00}
\right| \nonumber\\
 &&+ p_{\,4} \left| {11} \right\rangle \left\langle {11} \right| +
p_{\,5} \left| {02} \right\rangle \left\langle {02} \right| +
p_{\,6} \left| {12} \right\rangle \left\langle {12} \right| .
\end{eqnarray}

\noindent where $\sum\limits_i {p_{\,i} } = 1$.

We use convex programming method to determine the concrete form of
$p_i$. The basic idea is: by using the positive-definition or
semi-positive definition of partial transpose of separable state
$\tilde\sigma^*$ and Lagrangian multiplier limitation method to seek
for the solution of $p_i$, and then, we find out the closest
separable state for this class of state. The separable criterion is
a necessary and significant condition for $2\otimes 2$ and $2
\otimes 3$ quantum systems, so there are much more limitations in
the research field. If the value $p_i$ makes $\tilde\sigma^*=\sum
p_i|\psi_i\rangle\langle\psi_i|$ nonseparable, then this method is
invalid.

By calculation, the closest separable state of $\tilde {\rho }$ can
be expressed as£º
\begin{eqnarray}
\tilde {\sigma }^ * &=& \alpha \left( {\left| {02} \right\rangle
\left\langle {02} \right| + \left| {12} \right\rangle \left\langle
{12} \right|} \right)\; + \frac{3\beta + \gamma }{2}\left| {\psi ^ -
} \right\rangle
\left\langle {\psi ^ - } \right| \nonumber\\
&& + \frac{3\beta + \gamma }{6}\left( {\left| {\phi ^ + }
\right\rangle \left\langle {\phi ^ + } \right| + \left| {\phi ^ - }
\right\rangle \left\langle {\phi ^ - } \right| + \left| {\psi ^ + }
\right\rangle \left\langle {\psi ^ + } \right|} \right) .
 \end{eqnarray}

\noindent We need to point out that the detail process was given in
our classmate' unpublished thesis [36] that is provided in the
Appendix A.

Then the RE of $\tilde {\rho }$ is
\begin{eqnarray}
 E_{re}(\tilde\rho)&=&S\left( {\tilde {\rho }\left\| {\tilde {\sigma }^ * } \right.} \right) =
{\rm tr}\left( {\tilde\rho \log \tilde\rho } \right) - {\rm
tr}\left( {\tilde {\rho }\log
\tilde {\sigma }^ * } \right) \nonumber\\
 &=& 3\beta \log \left[ {\frac{6\beta }{3\beta +
\gamma }} \right] + \gamma \log \left[ {\frac{2\gamma }{3\beta +
\gamma }}
\right] \nonumber\\
 &=& \left( {1 - 2\alpha } \right)\log \left[
{\frac{2\left( {1 - \gamma - 2\alpha } \right)}{1 - 2\alpha }}
\right] + \gamma \log \left[ {\frac{\gamma }{1 - \gamma - 2\alpha }}
\right] .
\end{eqnarray}

Now, let us consider complex case {\it i.e.} two-parameter class of
states in $2 \otimes n$ quantum system [34] for $n \geq 3$, which
can be obtained from an arbitrary state in $2 \otimes n$ quantum
system by LOCC and are invariant under all unitary operations with
the form $U \otimes U$ on $2 \otimes n$ quantum system.
\begin{eqnarray}
 \rho &=& \alpha \sum\limits_{i = 0}^1 \sum\limits_{j = 2}^{n - 1} {\left|
{i\,j} \right\rangle \left\langle {i\,j} \right|} + \beta \left(
{\left| {00} \right\rangle \left\langle {00} \right| + \left| {11}
\right\rangle
\left\langle {11} \right|} \right) \nonumber\\
&&+ \frac{\beta + \gamma }{2}\left( {\left| {01} \right\rangle
\left\langle {01} \right| + \left| {10} \right\rangle \left\langle
{10} \right|} \right) + \frac{\beta - \gamma }{2}\left( {\left| {01}
\right\rangle \left\langle {10} \right| + \left\langle {10}
\right|\left\langle {01} \right|} \right) .
\end{eqnarray}

\noindent where $\{|ij\rangle: i=0,1,j=0,1, \cdots, n-1\}$ is an
orthonormal basis for $2 \otimes n$ quantum system, and the
coefficients satisfy the relation:
\begin{equation}
2(n-2)\alpha+3\beta+\gamma=1 .
\end{equation}
Remark when $\alpha=0$, this state equals the Werner state in $2
\otimes 2$ quantum system for $0\leq \gamma \leq1$. The state $\rho$
is entangled and distillable if and only if $\frac{1}{2}< \gamma \le
1$. We guess its closest separable state is£º
\begin{eqnarray}
 \sigma ^ * &=& \alpha \sum\limits_{i = 0}^1 \sum\limits_{j = 2}^{n - 1} {\left|
{i\,j} \right\rangle \left\langle {i\,j} \right|} \; + \frac{3\beta
+ \gamma }{6}\left( {\left| {\phi ^ + } \right\rangle \left\langle
{\phi ^ + } \right| + \left| {\phi ^ - } \right\rangle \left\langle
{\phi ^ - } \right| + \left| {\psi ^ + } \right\rangle \left\langle
{\psi ^ + } \right|} \right)\nonumber
\\
 &&+ \frac{3\beta + \gamma }{2}\left| {\psi ^ - } \right\rangle
\left\langle {\psi ^ - } \right| ,
\end{eqnarray}

\noindent The above expression is indeed the closest separable state
under the RE for the two-parameter class of states in $2 \otimes n$
quantum system which has been analyzed in our previous work [36].
(see Appendix B.)

The RE of  two parameter state is \fl \begin{eqnarray}
E_{re}(\rho)&=& S\left( {\rho \left\| {\sigma ^ * } \right.}
\right) = {\rm tr}\left(
{\rho \log \rho } \right) - {\rm tr}\left( {\rho \log \sigma ^ * } \right) \nonumber\\
&=& \left( {1 - 2\left( {n - 2} \right)\alpha } \right)\log \left[
{\frac{2\left( {1 - \gamma - 2\left( {n - 2} \right)\alpha }
\right)}{1 - 2\left( {n - 2} \right)\alpha }} \right] + \gamma \log
\left[ {\frac{\gamma }{1 - \gamma - 2\left( {n - 2} \right)\alpha
}}\right] \nonumber \\
&=& 3\beta \log \left( {\frac{6\beta }{3\beta + \gamma }} \right) +
\gamma \log \left( {\frac{2\gamma }{3\beta + \gamma }} \right)
\end{eqnarray}

\noindent where
\begin{eqnarray}
 &&2\left( {n - 2} \right)\alpha + 3\beta + \gamma = 1, \nonumber\\
 &&0 \le \alpha \le 1 / \left( {2n - 4} \right), \nonumber\\
 &&\frac{1}{2} \le \gamma \le 1 .
\end{eqnarray}

\noindent we draw the three dimensional picture of RE in fig.4.

\begin{figure}%[htbp]
\centering
\includegraphics[scale=0.86]{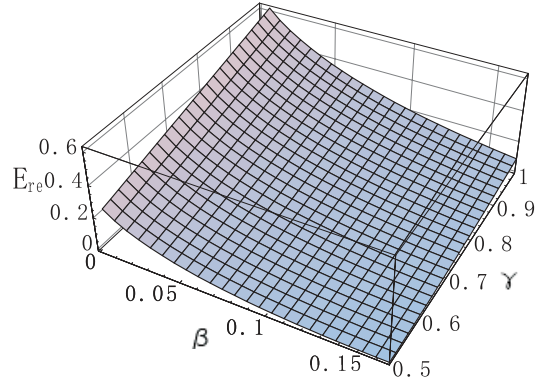}\caption{The relative
entropy of entanglement(RE) of two-parameter class of states in
$2\otimes n$ quantum system}
\end{figure}

Now we begin to calculate the analytical expression of the RGME for
two-parameter class of states in $2 \otimes n$ quantum system. Since
we obtain the expression of the RE, we want to ask whether the
closest separable state under the RE is also the closest separable
state under the RGME? The answer is yes for this state. We present a
proof in the following context.

From the definition of the RGME, we know the closest separable
state under the RGME is that under the Fidelity. To prove the
closest disentangled state to $\rho$ under the Fidelity metric is
$\sigma^*$, that is Eq.(54), we consider a slight variation around
$\sigma^*$ of the form $\sigma_\lambda = \left( {1 - \lambda }
\right)\sigma^* + \lambda \sigma $£¬ where $\sigma$ is any
separable state, then we just need to prove
\begin{equation}
\frac{d}{d\lambda }{\rm tr}\left\{ {\sqrt {\sqrt \rho \sigma
_\lambda \sqrt \rho } } \right\} \le 0 .
\end{equation}

\noindent Proof:
\begin{equation}
\sqrt \rho \sigma _\lambda \sqrt \rho = \rho ^{\frac{1}{2}}\left[
{\left( {1 - \lambda } \right)\sigma^* + \lambda \sigma} \right]\rho
^{\frac{1}{2}} = \left( {1 - \lambda } \right)\rho
^{\frac{1}{2}}\sigma^* \rho ^{\frac{1}{2}} + \lambda \rho
^{\frac{1}{2}}\sigma \rho ^{\frac{1}{2}} ,
\end{equation}

\noindent By choosing the appropriate basis sequence, $\sigma ^
* $ and $\rho $ can be expressed as the following matrix form
\begin{equation}
 \sigma ^ * = \left( {{\begin{array}{*{20}c}
 {A_{\,1} }  & 0  \\
 0  & {A_{\,2} }  \\
\end{array} }} \right)
, \rho = \left( {{\begin{array}{*{20}c}
 {B_{\,1} }  & 0  \\
 0  & {B_{\,2} }  \\
\end{array} }} \right).
\end{equation}

\noindent where
\[
A_{\,1} = \left( {{\begin{array}{*{20}c}
 {\displaystyle\frac{3\beta + \gamma }{6}}  & 0  & 0  & 0  \\
 0  & {\displaystyle\frac{3\beta + \gamma }{3}}  & { - \displaystyle\frac{3\beta + \gamma
}{6}}  & 0  \\
 0  & { - \displaystyle\frac{3\beta + \gamma }{6}}  & {\displaystyle\frac{3\beta + \gamma
}{3}}  & 0  \\
 0  & 0  & 0  & {\displaystyle\frac{3\beta + \gamma }{6}} \\
\end{array} }} \right),
\]

\[
B_{\,1} = \left( {{\begin{array}{cccc}
 \beta  & 0  & 0  & 0  \\[6pt]
 0  & {\displaystyle\frac{\beta + \gamma }{2}}  & {\displaystyle\frac{\beta - \gamma }{2}}
 & 0  \\[6pt]
 0 & {\displaystyle\frac{\beta - \gamma }{2}}  & {\displaystyle\frac{\beta + \gamma }{2}}
 & 0  \\[6pt]
 0  & 0  & 0 & \beta  \\[6pt]
\end{array} }} \right),
\]

\begin{equation}
 A_{\,2} = \left( {{\begin{array}{*{20}c}
 \alpha  & \ldots  & 0  \\
 \vdots  & \ddots  & \vdots  \\
 0  & \cdots  & \alpha  \\
\end{array} }} \right)£¬
, B_{\,2} = \left( {{\begin{array}{*{20}c}
 \alpha  & \ldots  & 0  \\
 \vdots  & \ddots & \vdots  \\
 0  & \cdots  & \alpha  \\
\end{array} }} \right).
\end{equation}

\noindent In addition, we can write the any separable state $\sigma$
as diagonal matrix block form
\begin{equation}
 \sigma = \left( {{\begin{array}{*{20}c}
 {X_{\,1} }  & 0  \\
 0  & {X_{\,2} }  \\
\end{array} }} \right).
\end{equation}

\noindent then we can deduce
\begin{eqnarray}
\fl \sqrt{ (1 - \lambda) \rho ^{\frac{1}{2}}\sigma^* \rho
^{\frac{1}{2}} + \lambda \rho ^{\frac{1}{2}}\sigma \rho
^{\frac{1}{2}} }&=& \sqrt{1 - \lambda}\left(\begin{array}{*{20}c}
{(B_1^\frac{1}{2} A_1 B_1^\frac{1}{2})}^\frac{1}{2} & 0  \\
0 & {(B_2^\frac{1}{2} A_2 B_2^\frac{1}{2})}^\frac{1}{2} \\
\end{array}  \right) \nonumber \\
&& + \sqrt{\lambda} \left(\begin{array}{cc} {(B_1^\frac{1}{2} X_1 B_1^\frac{1}{2})}^\frac{1}{2} & 0\\
0 & {(B_2^\frac{1}{2} X_2 B_2^\frac{1}{2})}^\frac{1}{2} \\
\end{array}  \right).
\end{eqnarray}

\noindent we have
\begin{eqnarray}
\frac{d}{d\lambda}{\rm tr}  {\sqrt { \sqrt \rho \sigma _\lambda
\sqrt \rho } } &=& \frac{d}{d\lambda} \left( \sqrt{1-\lambda} [F_{A_1}+F_{A_2}] +\sqrt{\lambda} [F_{X_1}+F_{X_2}] \right )  \nonumber  \\
&=& -\frac{1}{2}(1-\lambda)^{-\frac{1}{2}} (F_{A_1}+F_{A_2})
+\frac{1}{2}\lambda^{-\frac{1}{2}} (F_{X_1}+F_{X_2}).
\end{eqnarray}

\noindent where $F_{A_1}={\rm tr}\sqrt{B_1^\frac{1}{2} A_1
B_1^\frac{1}{2} }, F_{A_2}={\rm tr}\sqrt{B_2^\frac{1}{2} A_2
B_2^\frac{1}{2}}$ and $F_{X_1}={\rm tr}\sqrt{B_1^\frac{1}{2} X_1
B_1^\frac{1}{2}}, F_{X_2}={\rm tr}\sqrt{B_2^\frac{1}{2} X_2
B_2^\frac{1}{2}} $, respectively. We put $\lambda = 0$ into the
above expression, we get
\begin{equation}
\frac{d}{d\lambda }{\rm tr}\left\{ {\sqrt {\sqrt \rho \sigma
_\lambda \sqrt \rho } } \right\} = - \frac{1}{2}(F_{A_1}+F_{A_2}) .
\end{equation}

\noindent Because the Fidelity always larger than or equal to 0.
Finally we obtain the relation

\begin{equation}
\frac{d}{d\lambda }{\rm tr}\left\{ {\sqrt {\sqrt \rho \sigma
_\lambda \sqrt \rho } } \right\} \le 0 .
\end{equation}

\noindent the proof comes to an end.

The RGME for the two-parameter class of states in $ 2 \otimes n$
quantum system is thus
\begin{eqnarray}
\fl ~~~~~~~~\widetilde {E}_{\sin ^2} = 1 - F_n^2 &&= 1 - \left(
2\left( {n - 2} \right)\alpha + \frac{3\sqrt {\beta (3\beta + \gamma
)} }{\sqrt 6 } + \frac{\sqrt {3\beta \gamma + \gamma ^2} }{\sqrt 2 } \right)^2 \nonumber \\
&&= 1 - \left( {1 - \gamma - 3\beta + \frac{3\sqrt {\beta (3\beta +
\gamma )} }{\sqrt 6 } + \frac{\sqrt {\gamma(3\beta + \gamma) }
}{\sqrt 2 }} \right)^2.
\end{eqnarray}

\noindent where
\begin{equation}  0 \le \alpha \le \frac{1}{2n - 4},\quad -
\frac{1}{3} \le \beta \le \frac{1}{6},\quad \frac{1}{2} \le \gamma
\le 1.\end{equation}

\noindent The concrete proof process is given in Appendix C.

In virtue of the results of previous works [34,36], we know the
Negativity of this class of state is£º
\begin{eqnarray}
 N &=& \left( {2\left( {n - 2} \right)\alpha + 3\left| {\frac{\beta + \gamma
}{2}} \right| + \frac{1}{2}\left| {3\beta - \gamma } \right|} \right) - 1 \nonumber\\
&=& - 3\beta - \gamma + 3\left| {\frac{\beta + \gamma }{2}} \right|
+ \frac{1}{2}\left| {3\beta - \gamma } \right| .
\end{eqnarray}

\noindent Of course, the correspond three-dimension picture of the
RGME and the Negativity can be drawn in fig.4.

\begin{figure}[h]%[tbp]
\begin{center}
\includegraphics[scale=0.41]{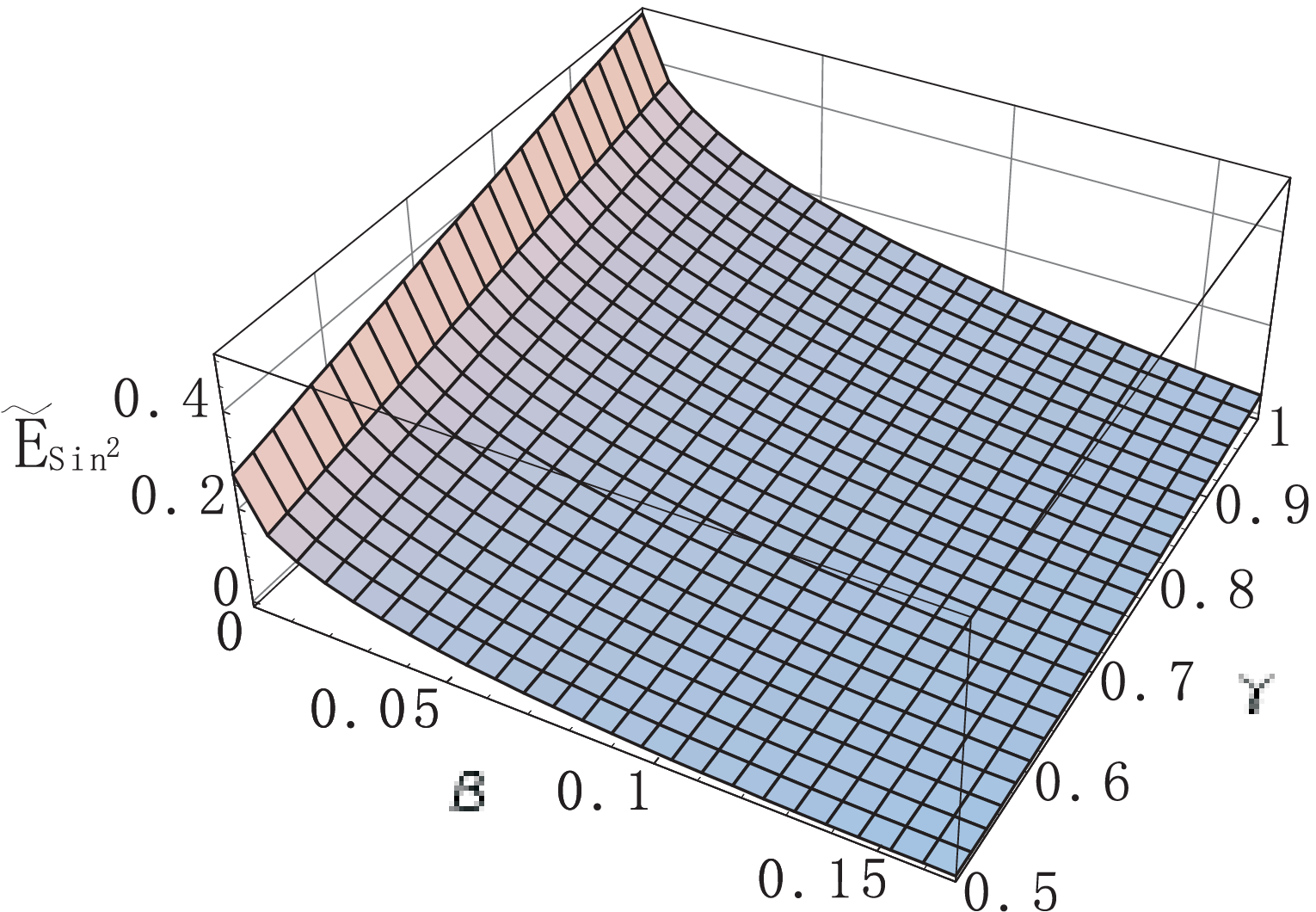}
\includegraphics[scale=0.39]{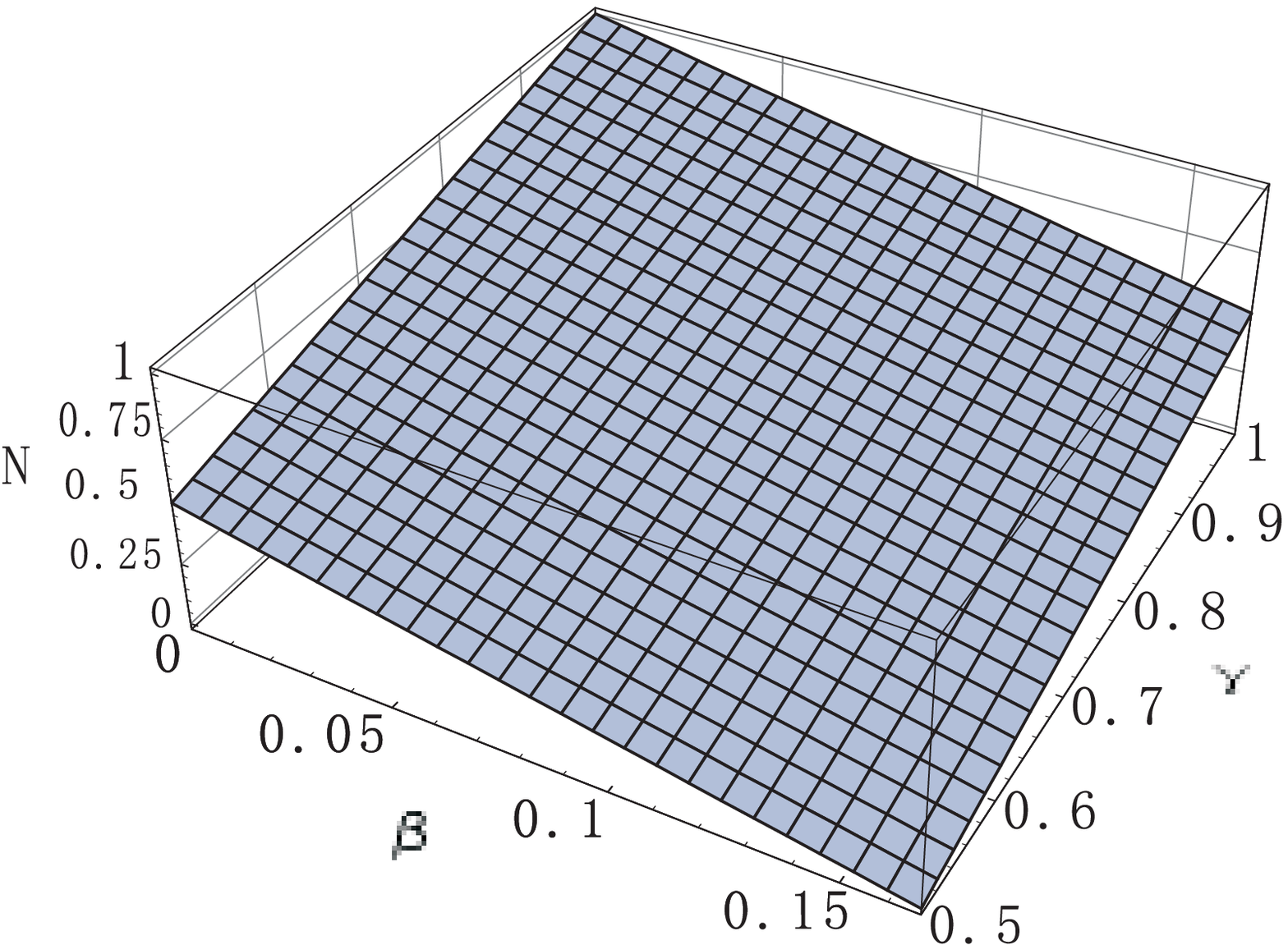}
\end{center}
\caption{Revised geometric measure of entanglement (RGME) and
Negativity for the two-parameter class of states in $2 \otimes n$
quantum system, respectively. }
\end{figure}

We can know the relation of three measures of entanglement for the
two-parameter class of states in $2 \otimes n$ quantum system by
drawing fig.5, {\it i.e.} $\widetilde {E}_{\sin ^2} \leq
E_{re}\leq Negativity$.

\begin{figure}%[htbp]
\centering
\includegraphics[scale=0.5]{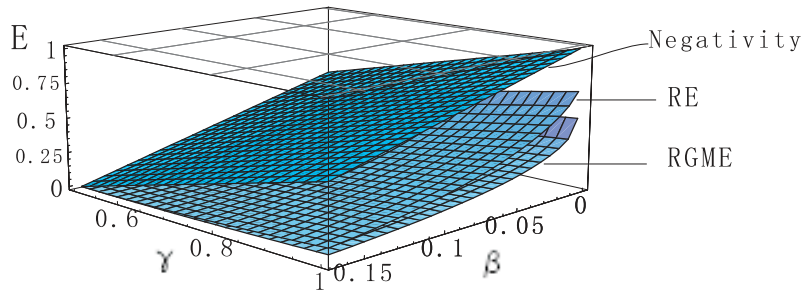}\caption{RGME, RE,
Negativity for the two-parameter class of states in $2 \otimes n$
quantum system. There is a relation $\widetilde {E}_{\sin ^2} \leq
E_{re}\leq Negativity$ clearly.}
\end{figure}

Ref.[37] gives the tight upper bound of EOF for the $2 \otimes n $
quantum system: $E_{f} \leq $Negativity.  And Ref.[38] presents a
lower bound for EOF on $2 \otimes n$ system and compares this
lower bound with the RE: $ E_{re} \leq E_{f} $. Besides, we know
any $2 \otimes n$ quantum state can be transformed to
two-parameter class of states Eq.(52) by LOCC. Based on the
requirement that the measure of entanglement is not increase under
LOCC, we acquire the relation for the two-parameter class of
states in $2 \otimes n$ quantum system:
\begin{equation}
\widetilde {E}_{\sin ^2}\leq E_{re}\leq E_{f} \leq Negativity .
\end{equation}

From above analysis and comparison, we see that using the RGME to
measure the entanglement is more appropriate for the two-parameter
class of states in $2 \otimes n$ quantum system.

\subsection{RGME of maximally entangled mixed state}
Ishizaka and Hiroshima first introduce the concept of the maximally
entangled mixed state [39] for which no more entanglement can be
created by global unitary operation, that is, acting on the system
as a whole.

In theory, by investigating the maximally entangled mixed state, we
can know the bounds on how the degree of mixing of a state limits
its entanglement. In practice, the mixture of the density matrix is
inevitably increased by the coupling between the quantum system and
its surrounding environment in all realistic systems. Therefore, it
is extremely important to understand the nature of entanglement for
general mixed states between two extremes of pure states and a
maximally mixed state.

Here, we extend results of two-parameter class of states in $2
\otimes n \left( {n> 2} \right)$ quantum system to the maximally
entangled mixed state [40] of two particles high dimensional
situation. Let the eigenvalue decomposition of $\rho$, i.e Eq.(52)
be
\begin{equation}
\rho=\Phi \Lambda \Phi^\dag ,
\end{equation}
where the eigenvalues ${\lambda_i}$ are sorted in nonascending
order. Obviously, we are capable of obtaining a mixed state $\rho
^{\prime}$ which achieves maximal Negativity by applying the
following global unitary transformation.
\begin{equation}
 U = \left( {U_1 \otimes U_2 } \right)\left(
\begin{array}{*{20}c}
 0  & \cdots  & 0  & 0  & \cdots  & 0  &0  & 1  \\
 {1 / \sqrt 2 }  & \cdots  & 0  & 0  & \cdots
& 0  & {1 / \sqrt 2 }  & 0  \\
 0  & \cdots  & 1  & 0  & \cdots  & 0  &0 & 0  \\
 \vdots  & \vdots  & \vdots  & \vdots  & \cdots
 & \vdots  & \vdots  & \vdots  \\
 { - 1 / \sqrt 2 }  & \cdots  & 0  & 0  & \cdots
&0  & {1 / \sqrt 2 }  &0  \\
 0  & \cdots  & 0  & 0  & \cdots  & 1  &0  &0  \\
 \vdots  & \vdots  & \vdots  & 0  & \cdots  &
\vdots  & \vdots  & \vdots  \\
 0  & 0  & 0  & 1  & 0  & 0  & 0& 0 \\
\end{array}  \right)D_\phi \Phi ^\dag ,
\end{equation}
where $U_1,U_2$ are two subsystem local unitary operations,
respectively. $D_\phi$ is a unitary diagonal matrix. The transformed
state is
\begin{eqnarray}
%\begin{array}{l}
 \rho ^\prime =U\rho U^\dag &=& \lambda _4 \left| {00} \right\rangle \left\langle {00} \right| +
\frac{\lambda _1 }{2}\left( {\left| {01} \right\rangle \left\langle
{01} \right| - \left| {01} \right\rangle \left\langle {10} \right| -
\left| {10} \right\rangle \left\langle {01} \right| + \left| {10}
\right\rangle
\left\langle {10} \right|} \right) \nonumber\\
&&+ \lambda _3 \left| {0\left( {n - 1} \right)} \right\rangle
\left\langle {0\left( {n - 1} \right)} \right| + \lambda _2 \left|
{1\left( {n - 1} \right)} \right\rangle \left\langle {1\left( {n -
1} \right)}
\right| \nonumber\\
 &=& \lambda _4 \left| {00} \right\rangle \left\langle {00} \right|
+ \lambda _1 \left| {\psi ^ - } \right\rangle \left\langle {\psi ^ -
} \right|
 + \lambda _3 \left| {0\left( {n - 1} \right)} \right\rangle
\left\langle {0\left( {n - 1} \right)} \right|\nonumber\\
&&+\lambda _2 \left| {1\left( {n - 1} \right)} \right\rangle
\left\langle {1\left( {n - 1} \right)} \right| ,
 %\end{array}
\end{eqnarray}

\noindent where $\lambda _i (i = 1,\,2,\,3,\,4)$ are four
eigenvalues which satisfy
\begin{equation}
\label{eq11} \lambda _1 + \lambda _2 + \lambda _3 + \lambda _4 = 1 .
\end{equation}

\noindent  the Negativity of $\rho ^{\prime}$ is same as the
count-part of 2-qubit situation [40].  Negativity of this state is
also the maximum.
\begin{equation}
N \left( {\rho ^{\prime}} \right) = \max (0,\sqrt {(\lambda _1 -
\lambda _3 )^2 + (\lambda _2 - \lambda _4 )^2} - \lambda _2 -
\lambda _4 ) .
\end{equation}

\noindent The closest separable state of $\rho ^{\prime}$ has the
following form:
\begin{eqnarray}
\sigma^\prime &=& p_{\,4} \left| {00} \right\rangle \left\langle
{00} \right| + p_{\,5} \left| {11} \right\rangle \left\langle {11}
\right| +
p_{\,1} \left| {\psi ^ - } \right\rangle \left\langle {\psi ^ - } \right| \nonumber\\
 && + p_3 \left| {0\left( {n - 1} \right)} \right\rangle
\left\langle {0\left( {n - 1} \right)} \right| + p_2 \left| {1\left(
{n - 1} \right)} \right\rangle \left\langle {1\left( {n - 1}
\right)} \right| ,
\end{eqnarray}

\noindent By convex programming method, we get the concrete form of
state $ \sigma^\prime$,
\begin{eqnarray}
 \sigma^\prime &=& \frac{\left( {\lambda _{\,1} + 2\lambda _4 }
\right)^2}{4\left( {\lambda _1 + \lambda _4 } \right)}\left| {00}
\right\rangle \left\langle {00} \right| + \frac{\lambda _1
^2}{4\left( {\lambda _1 + \lambda _4 } \right)}\left| {11}
\right\rangle \left\langle
{11} \right| \nonumber\\
&& + \frac{\lambda _1 \left( {\lambda _{\,1} + 2\lambda _4 }
\right)}{4\left( {\lambda _1 + \lambda _4 } \right)}\left( {\left|
{01} \right\rangle \left\langle {01} \right| + \left| {10}
\right\rangle \left\langle {10} \right| - \left| {01} \right\rangle
\left\langle {10}
\right| - \left| {10} \right\rangle \left\langle {01} \right|} \right) \nonumber\\
 && + \lambda _3 \left| {0\left( {n - 1} \right)} \right\rangle
\left\langle {0\left( {n - 1} \right)} \right| + \lambda _2 \left|
{1\left( {n - 1} \right)} \right\rangle \left\langle {1\left( {n -
1} \right)} \right| ,
\end{eqnarray}

\noindent the RE of $\rho^{\prime}$ is
\begin{equation}
\label{eq14} E_{re} \left( {\rho ^{\prime}} \right) = \lambda _{\,1}
\log \frac{2\left( {\lambda _{\,1} + \lambda _{\,4} }
\right)}{\left( {\lambda _{\,1} + 2\lambda _{\,4} } \right)} +
\lambda _{\,4} \log \frac{4\lambda _{\,4} \left( {\lambda _{\,1} +
\lambda _{\,4} } \right)}{\left( {\lambda _{\,1} + 2\lambda _{\,4} }
\right)^2} .
\end{equation}

\noindent It is easy to testify that the closest separable state
under the RE is same as that under the Fidelity by the same method
for the two-parameter class of states in $2 \otimes n$ quantum
system.

We can prove the Fidelity of the maximally entangled mixed state
$\rho ^{\prime}$ is
\begin{equation}
F = \lambda _2 + \lambda _3 + \frac{\left( {\lambda _1 + 2\lambda _4
} \right)}{2}\sqrt {\frac{\lambda _4 }{\lambda _1 + \lambda _4 }} +
\lambda _1 \sqrt {\frac{\lambda _1 + 2\lambda _4 }{2\left( {\lambda
_1 + \lambda _4 } \right)}} ,
\end{equation}
\noindent Above formula of Fidelity is without reference to $n$.
(see Appendix D). Then the RGME is£º
\begin{equation}
\widetilde {E}_{\sin ^2} =  1 - \left( {1-\lambda _1 - \lambda _4 +
\frac{\left( {\lambda _1 + 2\lambda _4 } \right)}{2}\sqrt
{\frac{\lambda _4 }{\lambda _1 + \lambda _4 }} + \lambda _1 \sqrt
{\frac{\lambda _1 + 2\lambda _4 }{2\left( {\lambda _1 + \lambda _4 }
\right)}} } \right)^2 .
\end{equation}

\noindent where $\lambda _1 + \lambda _2 + \lambda _3 + \lambda _4
= 1$. By comparing the amount of entanglement for different
entanglement measures Eq.(77) and Eq.(79), we find out that the RE
is an upper bound on the RGME for this special state, {\it i.e.}
\begin{equation}
\widetilde {E}_{\sin ^2}(\rho) \leq E_{re}(\rho).
\end{equation}
which also shows we use the RGME to measure the entanglement is more
appropriate for this class of state system.

\subsection{RGME of isotropic states}
Since the isotropic states are put forward, the properties of
isotropic states have been investigated and it has many
applications in different fields [44,45,46]. The state is called
isotropic because it is invariant under any $U_A\otimes U_B^*$
transformation
\begin{equation}
(U_A\otimes U_B^*)\rho_\alpha(U_A\otimes
U_B^*)^\dag=\rho_\alpha\mathcal{} .
\end{equation}
where $U$ is a unitary operator and $U^*$ is its conjugate [45]. In
essence, the isotropic states are a class of mixed states which are
convex mixtures of the maximally mixed state, $I_{d^2}=(I\otimes
I)/d^2$, with a maximally entangled state $\left| {\phi ^ + }
\right\rangle = \frac{1}{\sqrt d }\sum\limits_{i = 0}^{d - 1}
{\left| {ii} \right\rangle }$.

Here we present some new results about measure of entanglement for
the isotropic states including multi-particle and high dimension
generalization. Simultaneously, we review some other measures of
entanglement about the isotropic states.

First of all, for all isotropic qubit state£º
\begin{eqnarray}
\rho _\alpha &=& \alpha \left| {\phi ^2_ + } \right\rangle
\left\langle {\phi
^2_ + } \right| + \frac{1 - \alpha }{4}I \nonumber \\
& =& \left( {{\begin{array}{*{20}c} {\displaystyle\frac{1 + \alpha
}{4}}  & 0  & 0  & {\displaystyle\frac{\alpha }{2}} \\[6pt] 0 &
{\displaystyle\frac{1 - \alpha }{4}}  & 0  & 0 \\[6pt] 0 & 0  &
{\displaystyle\frac{1 - \alpha }{4}}  & 0  \\[6pt]
{\displaystyle\frac{\alpha }{2}}  & 0  & 0  & {\displaystyle\frac{1
+ \alpha }{4}} \\[6pt]
\end{array} }} \right),
\end{eqnarray}

\noindent where $\left| {\phi ^2_ + } \right\rangle = \frac{1}{\sqrt
2 }\left( {\left| {00} \right\rangle + \left| {11} \right\rangle }
\right)$.

We know when $ - \frac{1}{3} \le \alpha \le \frac{1}{3}$, $\rho
_\alpha $ is a separable state; when $\frac{1}{3} < \alpha \le 1$,
$\rho _\alpha $ is an entangled state, so the closest separable
state is $\sigma = \rho _{\frac{1}{3}}$. The RGME is
\begin{equation}
\widetilde {E}_{\sin ^2} = 1 - F^2 = 1 - \left( {\frac{1}{2}\sqrt
{\frac{3\left( {1 - \alpha } \right)}{2}} + \frac{1}{2}\sqrt
{\frac{1 + 3\alpha }{2}} } \right)^2 .
\end{equation}

For isotropic qutrit entangled state:
\begin{eqnarray}
\rho _\alpha &=&  \alpha \left| {\phi ^3_ + } \right\rangle
\left\langle {\phi
^3_ + } \right| + \frac{1 - \alpha }{9}I \nonumber\\
 &=& \left( {{\begin{array}{*{20}c}
 {\frac{1 + 2\alpha }{9}}  & 0  & 0  & 0  &
{\frac{\alpha }{3}} & 0  & 0 & 0 &
{\frac{\alpha}{3}}  \\
 0  & {\frac{1 - \alpha }{9}}  & 0  & 0  & 0
& 0  & 0  & 0  & 0  \\
 0  & 0  & {\frac{1 - \alpha }{9}}  & 0  & 0
& 0  & 0  & 0 & 0  \\
 0  & 0  & 0  & {\frac{1 - \alpha }{9}} & 0
& 0  & 0  & 0  & 0  \\
 {\frac{\alpha }{3}}  & 0 & 0  & 0  & {\frac{1 +
2\alpha }{9}}  & 0  & 0  & 0  & {\frac{\alpha }{3}}
 \\
 0  & 0  & 0  & 0  & 0  & { \frac{1 - \alpha
}{9}}  & 0  & 0  & 0  \\
 0  & 0  & 0 & 0  & 0 & 0  & {\frac{1
- \alpha }{9}}  & 0  & 0  \\
 0  & 0  & 0  & 0  & 0  & 0  & 0
& {\frac{1 - \alpha }{9}} & 0  \\
 {\frac{\alpha }{3}}  & 0  & 0  & 0  & {\frac{\alpha
}{3}}  & 0  & 0  & 0  & {\frac{1 + 2\alpha }{9}}
 \\
\end{array} }} \right) ,
 \end{eqnarray}

\noindent where $\left| {\phi ^3_ + } \right\rangle = \frac{1}{\sqrt
3 }\left( {\left| {00} \right\rangle + \left| {11} \right\rangle +
\left| {22} \right\rangle } \right)$.

We know when $ - \frac{1}{8} \le \alpha \le \frac{1}{4}$, $\rho
_\alpha $ is a separable state; when $\frac{1}{4} < \alpha \le 1$,
$\rho _\alpha $ is an entangled state. The closest separable state
is $\sigma=\rho _{\frac{1}{4}}$. The RGME is£º
\begin{equation}
\widetilde {E}_{\sin ^2} = 1 - F^2 = 1 - \left( {\frac{1}{3}\sqrt
{\frac{16(1 - \alpha )}{3}} + \frac{1}{3}\sqrt {\frac{1 + 8\alpha
}{3}} } \right)^2.
\end{equation}

For isotropic qu-quartit entangled state:
\begin{equation}
 \rho _\alpha = \alpha \left| {\phi ^4_ + } \right\rangle \left\langle {\phi
^4_ + } \right| + \frac{1 - \alpha }{16}I ,
\end{equation}

\noindent where $\left| {\phi ^4_ + } \right\rangle =
\frac{1}{2}\left( {\left| {00} \right\rangle + \left| {11}
\right\rangle + \left| {22} \right\rangle + \left| {33}
\right\rangle } \right)$.

We know when $ - \frac{1}{15} \le \alpha \le \frac{1}{5}$, $\rho
_\alpha $ is a separable state; when $\frac{1}{5} < \alpha \le 1$,
$\rho _\alpha $ is an entangled state. The closest separable state
is $\sigma = \rho _{\frac{1}{5}} $. The RGME is £º
\begin{equation}
\widetilde {E}_{\sin ^2} = 1 - F^2 = 1 - \left( {\frac{1}{4}\sqrt
{\frac{45(1 - \alpha )}{4}} + \frac{1}{4}\sqrt {\frac{1 + 15\alpha
}{4}} } \right)^2 .
\end{equation}

For the $d\times d$ isotropic state:
\begin{equation}
\rho _\alpha = \frac{1}{d^2}\left( {1 + \frac{d}{2}\alpha \Gamma }
\right),
\end{equation}

\noindent where $\Gamma = \sum\limits_{i = 1}^{d^2 - 1} {c_i \gamma
^i \otimes } \gamma ^i , \quad c_i = \pm 1 , \quad \gamma ^i$ is
Gell-Mann matrix.

We know when $ - \frac{1}{d^2 + 1} \le \alpha \le \frac{1}{d + 1}$,
$\rho _\alpha $ is a separable state; when $\frac{1}{d + 1} < \alpha
\le 1$, $\rho _\alpha $ is an entangled state. The closest separable
state is
\begin{equation}
\sigma = \rho _{\frac{1}{d + 1}} = \frac{1}{d^2}\left( {1 +
\frac{d}{2\left( {d + 1} \right)}\Gamma } \right) ,
\end{equation}

\noindent It is easy to get the analytical expression of the
Fidelity£º
\begin{equation}
F = \frac{d^2 - 1}{d}\sqrt {\frac{(1 - \alpha )}{d\left( {d + 1}
\right)}} + \frac{1}{d}\sqrt {\frac{1 + \left( {d^2 - 1}
\right)\alpha }{d}},
\end{equation}

\noindent The RGME is£º
\begin{equation}
\widetilde {E}_{\sin ^2} = 1 - F^2 = 1 - \left( {\frac{d^2 -
1}{d}\sqrt {\frac{(1 - \alpha )}{d\left( {d + 1} \right)}} +
\frac{1}{d}\sqrt {\frac{1 + \left( {d^2 - 1} \right)\alpha }{d}} }
\right)^2 .
\end{equation}

\noindent Furthermore, the isotropic state can be expressed as
[11,31]:
\begin{equation}
\rho _{iso} = \frac{1 - F^\prime}{d^2 - 1}\left( {I - \left| {\phi ^
+ } \right\rangle \left\langle {\phi ^ + } \right|} \right) +
F^\prime\left| {\phi ^ + } \right\rangle \left\langle {\phi ^ + }
\right| ,
\end{equation}

\noindent where $\left| {\phi ^ + } \right\rangle = \frac{1}{\sqrt d
}\sum\limits_{i = 0}^{d - 1} {\left| {ii} \right\rangle }, \quad
F^\prime = \Tr \left( {\rho _{\rm iso} \left| {\phi ^ + }
\right\rangle \left\langle {\phi ^ + } \right|} \right)$ is also the
Fidelity, different from the Fidelity in the RGME.

Ref.[14] gives the expression of the GME
\begin{equation}
E_{\sin ^2} = 1 - \frac{1}{d}\left( {\sqrt {F^\prime} + \sqrt
{\left( {1 - F^\prime} \right)\left( {d - 1} \right)} } \right)^2 .
\end{equation}

\noindent We can testify when $d = 2$,  $F^\prime = \frac{1 +
3\alpha }{4}$, we get the GME which is equal to the RGME.
\begin{equation}
E_{\sin ^2} = 1 - \left( {\frac{1}{2}\sqrt {\frac{1 + 3\alpha }{2}}
+ \frac{1}{2}\sqrt {\frac{3\left( {1 - \alpha } \right)}{2}} }
\right)^2 = \widetilde {E}_{\sin ^2},
\end{equation}

\noindent when $d = 3$, $F^\prime = \frac{1 + 8\alpha }{9}$, the GME
is equal to the RGME.
\begin{equation}
E_{\sin ^2} = 1 - \left( {\frac{1}{3}\sqrt {\frac{1 + 8\alpha }{3}}
+ \frac{1}{3}\sqrt {\frac{16\left( {1 - \alpha } \right)}{3}} }
\right)^2 = \widetilde {E}_{\sin ^2},
\end{equation}

\noindent when $d = 4$, $F^\prime = \frac{1 + 15\alpha }{16}$, the
GME is equal to the RGME.
\begin{equation}
E_{\sin ^2} = 1 - \left( {\frac{1}{4}\sqrt {\frac{45(1 - \alpha
)}{4}} + \frac{1}{4}\sqrt {\frac{1 + 15\alpha }{4}} } \right)^2 =
\widetilde {E}_{\sin ^2} ,
\end{equation}

\noindent when the dimension is $d$,  $F^\prime = \frac{1 + \left(
{d^2 - 1} \right)\alpha }{d^2}$, GME and RGME are also equal.
\begin{equation}
E_{\sin ^2} = 1 - \left( {\frac{d^2 - 1}{d}\sqrt {\frac{(1 - \alpha
)}{d\left( {d + 1} \right)}} + \frac{1}{d}\sqrt {\frac{1 + \left(
{d^2 - 1} \right)\alpha }{d}} } \right)^2 = \widetilde {E}_{\sin ^2}
.
\end{equation}

These not only show the GME and the RGME are equal for the isotropic
states, but also show that our revision for the GME is reasonable.
The contour figure of the isotropic state is given in fig.7. The
RGME in the undertone field is greater than that in the dark field.

\begin{figure}%[htbp]
\centering
\includegraphics[scale=0.71]{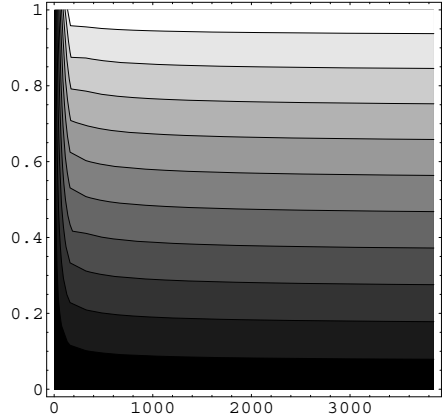}
\caption{the contour figure for the revised geometric measure of
entanglement (RGME) of the isotropic states, where undertone field
denotes higher density than counterpart in the dark field.}
\end{figure}

The concrete expression of RE [47] is £º
\begin{equation}
E_{re} = \log d + F^\prime\log F^\prime + \left( {1 - F^\prime}
\right)\log\frac{1 - F^\prime}{d - 1},
\end{equation}

\noindent where  $F^\prime = \frac{1 + \left( {d^2 - 1}
\right)\alpha }{d^2}$.

For this class of states, we still obtain the same relation
\begin{equation}
\widetilde {E}_{\sin ^2}(\rho) \leq E_{re}(\rho).
\end{equation}
like the special states in $2 \otimes n$ quantum system by drawing
figures. This result emphasizes the rationality of the revision and
it is an important conclusion through this paper simultaneously.

Other measures of entanglement about the isotropic state have been
obtained, concrete results are summarized in the following contexts.
Firstly, Concurrence and I-concurrence also have been given in
Ref.[47,48]. The Concurrence for the qubit isotropic state is
\begin{equation}
C = \left\{ \begin{array}{cc}
 0 \ \ &\ \ F^\prime < \frac{1}{2} ,\\
 2F^\prime - 1 \ \ & \ \ \displaystyle\frac{1}{2} \le F^\prime \le 1. \\
\end{array}  \right.
\end{equation}

\noindent qutrit isotropic state, I-concurrence(generalized
concurrence) is£º
\begin{equation}
C = \left\{ \begin{array}{cc}
 0\ \ &\ \ F^\prime < \frac{1}{3},\\
 \sqrt 3 \left( {F^\prime - \frac{1}{3}} \right)\ \ &\ \ \displaystyle\frac{1}{3} \le F^\prime \le 1.\\
\end{array}  \right.
\end{equation}

\noindent in turn, qudit isotropic state, I-concurrence is£º
\begin{equation}
C = \left\{ \begin{array}{cc}
 0\ \ &\ \  F^\prime < \frac{1}{d} ,\\
 \sqrt {\frac{2d}{d - 1}} \left( {F^\prime - \frac{1}{d}} \right)\ \ & \ \ \displaystyle\frac{1}{d}
\le F^\prime \le 1 .\\
\end{array}  \right.
\end{equation}

\noindent then the EOF for the isotropic state is £º
\begin{equation}
\fl E_f =\left\{ \begin{array}{cc}
 0,\ \ & \ \ F^\prime < \displaystyle\frac{1}{d}.\\
 h(x)+(1-x)\log(d-1), \ \ & \ \ \displaystyle\frac{1}{d} \le F^\prime <
 \frac{(d-1)(1-F^\prime)}{F^\prime} . \\
 d\log \displaystyle\frac{( {d - 1} )( {F^\prime - 1} )}{d - 2} + \log d ,\ \
 & \ \ \displaystyle \frac{4( {d - 1} )}{d^2}
 \le F^\prime \leq 1 .\\
\end{array}  \right.
\end{equation}

\noindent where $h\left( x \right) = - x\log x - \left( {1 - x}
\right)\log \left( {1 - x} \right)$, and $
x=\frac{F^\prime}{d}\left(1+\sqrt{\frac{(d-1)(1-F^\prime)}{F^\prime}}\right)^2$.
 Note when $d\rightarrow \propto$, we have $E_f\rightarrow F^\prime \log d$.

As for the relation about the Concurrence, the RGME and the RE, we
find their relation is uncertain by numerical analysis.

Now, we consider the generalized case, $n$-particle and $d$-level
isotropic state [23] is expressed as follows:
\begin{equation}
\rho \left( \alpha \right) = \left( {1 - \alpha}
\right)\frac{I}{d^n} + \alpha\left| {\psi ^ + } \right\rangle
\left\langle {\psi ^ + } \right|,\quad \quad \quad \quad 0 \le
\alpha \le 1.
\end{equation}

\noindent where $\left| {\psi ^ + } \right\rangle = \frac{1}{\sqrt d
}\sum\limits_{i = 1}^d {\left| {ii \cdots i} \right\rangle } $. The
closest separable state is£º
\begin{equation}
\rho \left( {\alpha_0 } \right) = \left( {1 - \alpha_0 }
\right)\frac{I}{d^n} + \alpha_0 \left| {\psi ^ + } \right\rangle
\left\langle {\psi ^ + } \right|, \quad\quad\quad\quad \alpha_0 =
\frac{1}{1 + d^{n - 1}}.
\end{equation}

\noindent when $n = 3, \quad d = 2$
\begin{equation}
F = 7\sqrt {\frac{1 - \alpha}{80}} + \sqrt {\frac{3\left( {1 +
7\alpha} \right)}{80}} ,
\end{equation}

\noindent when $n = 3,\quad d = 3 $
\begin{equation}
\quad F = 26\sqrt {\frac{1 - \alpha}{810}} + \sqrt {\frac{2\left( {1
+ 26\alpha} \right)}{405}} ,
\end{equation}

\noindent when $n = 4,\quad d = 2 $
\begin{equation}
F = 15\sqrt {\frac{1 - \alpha}{288}} + \sqrt {\frac{1 +
15\alpha}{96}} ,
\end{equation}

\noindent By the mathematical induction method, we get
\begin{equation}
F= \left( {d^n - 1} \right)\sqrt {\frac{1 - \alpha}{d^n\left( {d^n +
d} \right)}} + \sqrt {\frac{1 + \left( {d^n - 1}
\right)\alpha}{d^n}\frac{d + 1}{d^n + d}},
\end{equation}

\noindent notice when $n = 2$, we get the formula (84) of the
two-particle isotropic state again.

Finally, the explicit expression of the RGME for the generalized
case is
\begin{equation}
\widetilde {E}_{\sin ^2} = 1 - \left( {\left( {d^n - 1} \right)\sqrt
{\frac{1 - \alpha}{d^n\left( {d^n + d} \right)}} + \sqrt {\frac{1 +
\left( {d^n - 1} \right)\alpha}{d^n}\frac{d + 1}{d^n + d}} }
\right)^2 .
\end{equation}
when $n = 2$, we get the formula (90) of the two-particle
isotropic state again. These indicate our revision is reasonable,
too.

\subsection{RGME of some multi-particle bound entangled states}

Multiparticle entanglement exhibits a much richer structure than
biparticle entanglement, even in the simplest case, the
quantification of multiparticle entanglement is a hard computable
problem. It is thus worth seeking cases in which one can explicitly
obtain an expression to measure the amount of entanglement.

Bound multi-particle entangled states, the peculiar class of states
plays an important role in many calculations of entanglement
measure. Here, we determine analytically the entanglement in terms
of RGME for two multiparticle bound entangled states in Ref.[15] by
a purification procedure. The result shows that the RGME is equal to
the GME which elucidates the RGME is an appropriate measure of
entanglement comparing to other measures of entanglement, again. In
order to explain explicitly, we give a requisite theorem about
purification [30].

\noindent {\textbf{Uhlmann theorem:}} Assume $\rho,\sigma$ are
states of quantum system Q, introduce the second quantum system R
with dimension greater than or equal to the dimension of Q, then
\begin{equation}
F(\rho,\sigma)=max|\langle\varphi|\psi\rangle| .
\end{equation}
where the maximum runs over all purification $|\psi\rangle$ of
$\rho$, $|\varphi\rangle$ of $\sigma$ in RQ.

Firstly, we consider the Smolin's four-party unlockable bound
entangled state:
\begin{eqnarray}
 \rho ^{ABCD} &=& \frac{1}{4}\sum\limits_{i = 0}^3 {\left( {\left| {\psi _i }
\right\rangle \left\langle {\psi _i } \right|} \right)_{AB} \otimes
} \left( {\left| {\psi _i } \right\rangle \left\langle {\psi _i }
\right|}
\right)_{CD} \nonumber\\
& =& \frac{1}{4}\sum\limits_{i = 0}^3 {\left| {X_i } \right\rangle
\left\langle {X_i } \right|},
\end{eqnarray}

\noindent where
\begin{eqnarray}
 \left| {X_0 } \right\rangle = \frac{1}{\sqrt 2 }\left( {\left| {0000}
\right\rangle + \left| {1111} \right\rangle } \right)\nonumber ,\\
 \left| {X_1 } \right\rangle = \frac{1}{\sqrt 2 }\left( {\left| {0011}
\right\rangle + \left| {1100} \right\rangle } \right) \nonumber ,\\
 \left| {X_2 } \right\rangle = \frac{1}{\sqrt 2 }\left( {\left| {0101}
\right\rangle + \left| {1010} \right\rangle } \right) \nonumber,\\
 \left| {X_3 } \right\rangle = \frac{1}{\sqrt 2 }\left( {\left| {0110}
\right\rangle + \left| {1001} \right\rangle } \right).
 \end{eqnarray}

\noindent We can see $\rho ^{ABCD}$ has been written down in the
eigenvalue decomposition itself, that is
\begin{equation}\rho ^{ABCD} = \frac{1}{4}\sum\limits_{i = 0}^3 {\left| {X_i }
\right\rangle \left\langle {X_i } \right|} = \sum\limits_{i = 0}^3
{p_i \left| {X_i } \right\rangle \left\langle {X_i } \right|}  .
\end{equation}

Assume the closest separable state is $\sigma$, its eigenvalue
decomposition is $\sigma = \sum_{i=0}^3 q_i\left | \phi
\right\rangle \left\langle \phi \right|$, where $q_0 = 1, q_1 = q_2
=q_3 = 0$, $\left| \phi \right\rangle = \otimes _{_{i = 0} }^3
\left( {c_i \left| 0 \right\rangle + s_i \left| 1 \right\rangle }
\right)$, where $c_i\equiv\cos \theta_i, s_i\equiv\sin \theta_i$
with $0 \leq \theta_i \leq \frac{\pi}{2}$, so
\begin{eqnarray}
 & &\left| \psi \right\rangle = \sum\limits_{i = 0}^3 {\sqrt {p_i } \left| {X_i
} \right\rangle \left| {i^{R_A }} \right\rangle } ,\nonumber\\
 & &\left| \varphi \right\rangle = \sum\limits_{i = 0}^3 {\sqrt {q_i } \left|
\phi \right\rangle \left| {i^{R_B }} \right\rangle } = \left| \phi
\right\rangle \left| {i^{R_B }} \right\rangle .
\end{eqnarray}

\noindent Because arbitrariness of purification, we choose
$i^{R_A}=i^{R_B}$, then
\begin{eqnarray}
 F &=& \left\langle \varphi \right|\left. \psi \right\rangle=\left\langle
\phi \right|\sum\limits_{i = 0}^3 {\sqrt {p_i } \left| {X_i }
\right\rangle }= \sum\limits_{i = 0}^3 {\sqrt {p_i } \left\langle
\phi \right.\left| {X_i
} \right\rangle }\nonumber \\
&=& \sqrt {\frac{p_0 }{2}} \left( {c_1 c_2 c_3 c_4 + s_1 s_2 s_3 s_4
} \right) + \sqrt {\frac{p_1 }{2}} \left( {c_1 c_2 s_3 s_4 + s_1 s_2
c_3 c_4 }
\right) \nonumber\\
& &  + \sqrt {\frac{p_2 }{2}} \left( {c_1 s_2 c_3 s_4 + s_1 c_2 s_3
c_4 } \right) + \sqrt {\frac{p_3 }{2}} \left( {c_1 s_2 s_3 c_4 + s_1
c_2 c_3 s_4 } \right)
\end{eqnarray}

\noindent by Cauchy-Schwarz inequality $\left\langle v \right|\left.
v \right\rangle \left\langle w \right|\left. w \right\rangle \ge
\left| {\left\langle v \right|\left. w \right\rangle } \right|^2$,
we obtain
\begin{equation}
F = \sqrt {\frac{1}{2}} \;,\quad  \widetilde{E}_{\sin ^2}=
\frac{1}{2} .
\end{equation}

\noindent In addition, Ref.[15] conjectures its closest separable
mixed state is
\begin{eqnarray}
\sigma&=&\frac{1}{8}(|0000\rangle\langle0000|+|1111\rangle\langle1111|)+|0011\rangle\langle0011|
+|1100\rangle\langle1100|\nonumber\\
&&+|0101\rangle\langle0101|+|1010\rangle\langle1010|+|0110\rangle\langle0110|+|1001\rangle\langle1001|.
\end{eqnarray}

\noindent We compute the RGME using above suspected closet
separable state. The results are indeed $F=\frac{1}{\sqrt{2}}$,
$\widetilde {E}_{\sin ^2}=\frac{1}{2}$ which are same as Eq.(117).
Hence, we show the conjecture is valid from the inverted angle.

Next, we consider the Dur's $N$-party Bell-inequality-violating
bound entangled states ($N \geq 4$) [49]
\begin{equation}
\rho _N \left( x \right) = x\left| {\psi _G } \right\rangle
\left\langle {\psi _G } \right| + \frac{1 - x}{2N}\sum\limits_{k =
1}^N {\left( {P_k + \bar {P}_k } \right)} ,
\end{equation}

\noindent where
\begin{eqnarray}
& &\left| {\psi _G } \right\rangle = \displaystyle\frac{1}{\sqrt 2
}\left( {\left| {0^{ \otimes N}} \right\rangle + \left| {1^{ \otimes
N}} \right\rangle } \right),\nonumber\\
& &P_K = \left| {u_k } \right\rangle \left\langle {u_k }
\right|,\quad \quad \left| {u_k } \right\rangle = \left| 0
\right\rangle _1 \left| 0 \right\rangle _2 \cdots \left| 1
\right\rangle _k \cdots \left| 0
\right\rangle _N .\nonumber\\
& &\bar {P}_K = \left| {v_k } \right\rangle \left\langle {v_k }
\right|,\quad \quad \left| {v_k } \right\rangle = \left| 1
\right\rangle _1 \left| 1 \right\rangle _2 \cdots \left| 0
\right\rangle _k \cdots \left| 1 \right\rangle _N .
\end{eqnarray}

\noindent the eigenvalue decomposition can be conveniently written
as
\begin{eqnarray}
&& \rho _N \left( x \right) = \sum\limits_{i = 0}^{N - 1} {p_i
\left| {\xi _i } \right\rangle \left\langle {\xi _i } \right|} =
\left| {\psi \left( {x,\left\{ {q,r} \right\}} \right)}
\right\rangle \left\langle {\psi \left(
{x,\left\{ {q,r} \right\}} \right)} \right|, \nonumber\\
 &&\sigma = \left| \phi \right\rangle \left\langle \phi \right| .
\end{eqnarray}
where
\begin{equation}
\left| {\psi \left( {x,\left\{ {q,r} \right\}} \right)}
\right\rangle=\sqrt{x}| {\psi _G } \rangle +\sqrt{1-x}
\sum\limits_{k =
1}^{N}(\sqrt{q_{k}}|u_k\rangle+\sqrt{r_{k}}|v_k\rangle).
\end{equation}

\noindent Through the way of purification, we obtain£º
\begin{eqnarray}
 \left| \psi \right\rangle = \sum\limits_{i = 0}^{N - 1} {\sqrt {p_i }
\left| {\xi _i } \right\rangle \left| {i^{R_A }} \right\rangle }\nonumber, \\
 \left| \varphi \right\rangle = \sum\limits_{i = 0}^{N - 1} {\sqrt {q_i }
\left| \phi \right\rangle \left| {i^{R_B }} \right\rangle }.
 \end{eqnarray}

\noindent choose $i^{R_A}=i^{R_B}$, then we have
\begin{eqnarray} F
&=& \left\langle \varphi \right|\left. \psi \right\rangle =
\left\langle \varphi \right|\sum\limits_{i = 0}^{N - 1} {\sqrt {p_i
} \left| {\xi _i } \right\rangle } \\
& =& \sqrt {\frac{x}{2}} \left( {c_1 \cdots c_N + s_1 \cdots s_N }
\right) +\sqrt {1 - x} \sum\limits_{k = 1}^N {\left( {\sqrt {q_k }
c_1 \cdots s_k \cdots c_N + \sqrt {r_k } s_1 \cdots c_k \cdots s_N }
\right)}\nonumber,
\end{eqnarray}

\noindent similarly, using the Cauchy-Schwarz inequality, we obtain
\begin{equation}
F = \sqrt {\frac{2 - x}{2}},~~ \widetilde {E}_{\sin ^2} =
\frac{x}{2}.
\end{equation}

It is clear that the GME and the RGME are equal for above two bound
entangled states. Ref.[15] presents a conjecture concerning the
closest separable state of Dur's bound entangled state. The
suspected form is
\begin{equation}
\fl \rho_N(x)=x(| {0 \cdots 0}\rangle \langle {0 \cdots 0} | +| {1
\cdots 1} \rangle \langle {1 \cdots 1} |)+ \frac{1 -
x}{2N}\sum\limits_{k = 1}^N {\left( {P_k + \bar {P}_k } \right)},
\end{equation}

\noindent By Uhlmann theorem, we can easily obtain
\begin{eqnarray}
\fl |\psi\rangle=\sqrt{x}| {0 \cdots
0}\rangle|i^R\rangle+\sum\limits_{k = 1}^N \sqrt{\frac{1-x}{2N}}P_k
|i^R\rangle+ \sum\limits_{k = 1}^N \sqrt{\frac{1-x}{2N}}\bar
{P}_k|i^R\rangle+\sqrt{x}| {1 \cdots
1}\rangle|i^R\rangle,\nonumber\\
\fl |\varphi\rangle=\sqrt{\frac{x}{2}}| {0 \cdots
0}\rangle|i^R\rangle+\sum\limits_{k = 1}^N \sqrt{\frac{1-x}{2N}}P_k
|i^R\rangle+ \sum\limits_{k = 1}^N \sqrt{\frac{1-x}{2N}}\bar
{P}_k|i^R\rangle+\sqrt{\frac{x}{2}}| {1 \cdots 1}\rangle|i^R\rangle
.\nonumber\\
\end{eqnarray}

\noindent If the overlap satisfies
$\langle\varphi|\psi\rangle=\sqrt{1-\frac{x}{2}}$  then the
suspected separable state is valid. But the result of computation
doesn't satisfy this condition, so we say this conjecture is
invalid.

Thus, we have presented analytical results on how much entanglement
is bound in two distinct multipartite bound entanglement states
using the revised measure. For these states, the RE is still an
upper bound on the RGME. For example, the RE of the Smolin state is
$1$ [15] which is larger than its RGME $\frac{1}{2}$.

\section {Conclusion}\label {sec5}
The merit of this revised measure RGME lies on suiting for
any-partite system with any dimension. The revision of the GME
becomes more accurate. Because the RGME abandons the condition
that the closest separable state is pure state, even for the case
of the pure state, simultaneously uses the Fidelity to substitute
the overlap in view of the relation between the Fidelity and the
overlap, hence it can be expressed congruously. The essence of
problem is attributed to find out the closest separable state,
naturally, we need not use the convex hull construction to
consider the case of the mixed state. We have presented analytical
results about measure of entanglement of some special
multi-particle cases for which other measures of entanglement are
bigger than RGME, hence the advantage of RGME is exhibited
clearly.

Some properties of RGME are presented in the proposition form. We
discover the RGME is smaller than or equal to the EOF (or the ER)
in the bipartite pure state setting. For any pure state, the RGME
is smaller than or equal to the trace distance. Besides, we obtain
a relation between the von Neumann entropy and the trace distance.

The revised entanglement quantifier is used to quantify the
entanglement of some special states. We acquire two main bound
conditions, one is $\widetilde {E}_{\sin ^2} \left( \rho
\right)\leq E_{\sin ^2} \left( \rho \right) $ for the case of pure
state, another is $\widetilde {E}_{\sin ^2}\leq E_{re}\leq E_{f}
\leq Negativity $ for the two-parameter class of states in $2
\otimes n$ quantum system. The bound condition $\widetilde
{E}_{\sin ^2}\leq E_{re}$ is still valid for the bipartite
maximally entangled state, isotropic state, Smolin and Dur
multipartite bound states. From these conclusions, we see our RGME
is reasonable and has explicit application.

However, we should point out the disadvantage of RGME. Like the
RE, the search of the closest separable state is necessary for
calculation. In fact, this is a tough task. Certainly, GME and EOF
use the convex hull construction to deal with the case of mixed
state which is not easier than the former. Fortunately, for some
special states, the closest separable state under RE is also that
under RGME, which simplifies the difficulty greatly and makes the
calculation realizable. From this sense, we think this quantifier
outbalances other candidates of entanglement.

%Up to now, the search of the closest separable state is still a
%challenging problem.
In order to alleviate and overcome the difficulty of finding the
closest separable state, people provide many methods for
calculation of RE in the literature. The convex programming method
[35], the numerical value analysis method [8], etc, but they are
effective only for the special scenarios. The avail method which
suits to any state is to have a guess to what the minimum for a
pure state should be, then use the formal proof to testify, {\it
i.e.} considering the gradient, see Ref.[8].

Note Refs.[50,51] present an entanglement monotone derived from
Grover's algorithm called the Groverian entanglement. For a pure
state, the Groverian entanglement is equal to the GME, but their
physical meanings and springboards are completely different.
Because the Groverian entanglement is motivated by a quantum
algorithm, while the GME is motivated from a geometric viewpoint.
Groverian entanglement demonstrates how well a state performs an
input to Grover's search algorithm depends critically upon the
entanglement presented in that state; the more the entanglement,
the less well the algorithm performs. The GME is the sine of the
angle between the pure state and its closest separable state, the
stronger the entanglement of state becomes, the larger the angle
between them is. The Groverian entanglement is introduced just for
the pure state of multiple qubits, GME is suitable for
any-particle system with any dimension. On the basis of the
results about pure states, the Groverian entanglement is
generalized to the case of mixed states [52], but the operational
explanation can not be generalized to mixed states. In this paper
the GME is revised, while the RGME still maintains the inherent
advantages of the GME and has clear physical meaning [53]. It
happens that the forms of generalized Groverian entanglement [52]
and the RGME provided in this paper are coincident. It's worth
emphasizing that our work finished independently and in a
different way.

To the best of our knowledge, corresponding results have not been
obtained for other measures of entanglement. Recently, a connection
is identified between the GME and the entanglement witnesses [16],
which can in principle be measured locally. So, we can render the
GME experimentally verifiable. The connection between the
generalized robustness and the geometric measure of entanglement is
also presented in Ref.[54]. In view of these works, we wish to find
out the deeper relation between the RGME and other measures of
entanglement.

In conclusion, we believe our analysis is helpful for better
understanding the essence of amount of the entanglement. Because
many entangled quantifiers exist, it is important to explore their
relations. We believe, this should be a major goal in the theory
of entanglement and hope that the discussion in this paper can
give some help in this sense.

 \ack
We thank Ren-Gui Zhu, Si-Ping Liu, Feng Xu, Xiao-Shan Ma, Ning-Bo
Zhao and Xiao-Qiang Su for useful discussions. This project was
supported by the National Basic Research Programme of China under
Grant No 2001CB309310, the National Natural Science Foundation of
China under Grant No 60573008.

\section*{Appendix A}
\appendix
\setcounter{section}{1} In this section, we use the convex
programming method to find the closest separable state of the
two-parameter $2\otimes 3$ quantum system.
\begin{eqnarray}
 \tilde {\rho } &=& \beta \left| {\psi ^ + } \right\rangle \left\langle
{\psi ^ + } \right| + \gamma \left| {\psi ^ - } \right\rangle
\left\langle {\psi ^ - } \right| + \beta \left( {\left| {00}
\right\rangle \left\langle {00} \right| + \left| {11} \right\rangle
\left\langle {11} \right|} \right)\nonumber
\\
&&+ \alpha \left( {\left| {02} \right\rangle \left\langle {02}
\right| + \left| {12} \right\rangle \left\langle {12} \right|}
\right).
\end{eqnarray}

In light of the method in Ref.[35] the closest separable state of
$\tilde {\rho }$ has the following form£º
\begin{eqnarray}
 \tilde {\sigma }^ * &=& p_{\,1} \left| {\psi ^ + } \right\rangle \left\langle
{\psi ^ + } \right| + p_{\,2} \left| {\psi ^ - } \right\rangle
\left\langle {\psi ^ - } \right| + p_{\,3} \left| {00} \right\rangle
\left\langle {00}
\right|\nonumber \\
&& + p_{\,4} \left| {11} \right\rangle \left\langle {11} \right| +
p_{\,5} \left| {02} \right\rangle \left\langle {02} \right| +
p_{\,6} \left| {12} \right\rangle \left\langle {12} \right|.
\end{eqnarray}

\noindent where $\sum\limits_i {p_{\,i} } = 1$. Then the partial
transpose of $ \tilde {\sigma }^*$ is
\begin{eqnarray}
\left( {\tilde {\sigma }^ * } \right)^{T_B } &=& \frac{p_{\,1}
}{2}\left( {\left| {01} \right\rangle \left\langle {01} \right| +
\left| {00} \right\rangle \left\langle {11} \right| + \left| {11}
\right\rangle \left\langle {00} \right| + \left| {10} \right\rangle
\left\langle {10}
\right|} \right) \nonumber\\
&& + \frac{p_{\,2} }{2}\left( {\left| {01} \right\rangle
\left\langle {01} \right| - \left| {00} \right\rangle \left\langle
{11} \right| - \left| {11} \right\rangle \left\langle {00} \right| +
\left| {10}
\right\rangle \left\langle {10} \right|} \right) \nonumber \\
&&+ p_{\,3} \left| {00} \right\rangle \left\langle {00} \right| +
p_{\,4} \left| {11} \right\rangle \left\langle {11} \right| +
p_{\,5} \left| {02} \right\rangle \left\langle {02} \right| +
p_{\,6} \left| {12} \right\rangle \left\langle {12} \right|.
\end{eqnarray}

\noindent Because $\tilde {\sigma }^ * $ is separable state, we know
$\left( {\tilde {\sigma }^
* } \right)^{T_B }$ should be positive definite or semi-positive definite matrix according to the separable
criterion, so we obtain an inequality condition:
\begin{equation}
\label{eq16} p_{\,3} p_{\,4} - \left( {\frac{p_{\,1} - p_{\,2} }{2}}
\right)^2 \ge 0.
\end{equation}

The RE of  $\tilde {\rho }$ can be expressed as£º
\begin{eqnarray}
S\left( {\tilde {\rho }\left\| {\tilde {\sigma }^ * } \right.}
\right) &=& {\rm tr}\left( {\tilde {\rho }\log \tilde {\rho }}
\right) - {\rm tr}\left( {\tilde
{\rho }\left\| {\tilde {\sigma }^ * } \right.} \right) \nonumber\\
&=& {\rm tr}\left( {\tilde {\rho }\log \tilde {\rho }} \right) +
f\left( {p_{\,i} } \right) .
%\end{array}
\end{eqnarray}
\noindent Let
\begin{eqnarray}
F\left( {p_{\,i} } \right) &=& f\left( {p_{\,i} } \right) + \lambda
\left( {\sum\limits_i {p_{\,i} } - 1} \right) + \eta \left( {p_{\,3}
p_{\,4} -
\left( {\frac{p_{\,1} - p_{\,2} }{2}} \right)^2} \right) \nonumber\\
 &=& - \left( {\beta \log p_{\,1} + \gamma \log p_{\,2} +
\beta \log p_{\,3} + \beta \log p_{\,4} + \alpha \log p_{\,5} +
\alpha \log
p_{\,6} } \right) \nonumber\\
&& +\lambda \left( {\sum\limits_i {p_{\,i} } - 1} \right) + \eta
\left( {p_{\,3} p_{\,4} - \left( {\frac{p_{\,1} - p_{\,2} }{2}}
\right)^2} \right),
\end{eqnarray}

\noindent where $\lambda,\eta $ are Lagrange multipliers. The
problem comes to solve the extremum of $f\left( {p_{\,i} } \right)$
with a constraint, {\it i.e.} solve the following equations set.
\begin{equation}
\left\{ {\begin{array}{l}
 \displaystyle\frac{\partial F\left( {p_{\,i} } \right)}{\partial p_{\,j} } = 0,\;\quad
\quad \left( {j = 1,2,3,4} \right), \\
 \sum\limits_i {p_{\,i} } - 1 = 0,\quad \eta \left( {p_{\,3} p_{\,4} -
\left( {\displaystyle\frac{p_{\,1} - p_{\,2} }{2}} \right)^2}
\right) = 0,\quad \quad
\left( {\eta \ge 0} \right) \\
 \end{array}} \right.
\end{equation}

\noindent Through calculation we get two groups of solution. The
first group of solution is
\begin{eqnarray}
 & &p_{\,1} = p_{\,3} = p_{\,4} = \displaystyle\frac{3\beta + \gamma }{6},\nonumber \\
 & &p_{\,2} = \displaystyle\frac{3\beta + \gamma }{2}, \nonumber\\
 & &p_{\,5} = p_{\,6} = \alpha.
\end{eqnarray}

\noindent the second group of solution is
\begin{eqnarray}
 & &p_{\,1} ^\prime =  \displaystyle\frac{3\beta + \gamma }{2}, \nonumber\\
 & &p_{\,2} ^\prime =  \displaystyle\frac{\gamma \left( {3\beta + \gamma } \right)}{2\left(
{2\beta + \gamma } \right)}, \nonumber\\
 & &p_{\,3} ^\prime = p_{\,4} ^\prime =  \displaystyle\frac{\beta \left( {3\beta + \gamma }
\right)}{2\left( {2\beta + \gamma } \right)} \nonumber\\
 & &p_{\,5} ^\prime = p_{\,6} ^\prime = \alpha .
\end{eqnarray}

\noindent Due to the equation \begin{equation} \fl ~~~f\left(
{p_{\,i} } \right) = - \left( {\beta \log p_{\,1} + \gamma \log
p_{\,2} + \beta \log p_{\,3} + \beta \log p_{\,4} + \alpha \log
p_{\,5} + \alpha \log p_{\,6} } \right),\end{equation}

\noindent and the relations
\begin{equation}
1 / 2 \le \gamma \le 1, ~~~~\beta >0.
\end{equation}
\noindent we know
\begin{equation}
\label{eq22} f\left( {p_{\,i} } \right) \le f\left( {p_{\,i}
^\prime} \right).
\end{equation}

\noindent Evidently, we choose the first group of solution. At the
end, we obtain the closest separable state of $\tilde {\rho }$
\begin{eqnarray}
 \tilde {\sigma }^ * &=& \alpha \left( {\left| {02} \right\rangle \left\langle
{02} \right| + \left| {12} \right\rangle \left\langle {12} \right|}
\right)\; + \frac{3\beta + \gamma }{2}\left| {\psi ^ - }
\right\rangle
\left\langle {\psi ^ - } \right| \nonumber\\
 &&+ \frac{3\beta + \gamma }{6}\left( {\left| {\phi ^ + }
\right\rangle \left\langle {\phi ^ + } \right| + \left| {\phi ^ - }
\right\rangle \left\langle {\phi ^ - } \right| + \left| {\psi ^ + }
\right\rangle \left\langle {\psi ^ + } \right|} \right) .
 \end{eqnarray}

\section*{Appendix B}
\appendix
\setcounter{section}{2} In this appendix we show a proof that the
closest separable state of two-parameter class of stats in $2
\otimes n$ quantum system
\begin{eqnarray}
\rho &=& \alpha \sum\limits_{i = 0}^1 \sum\limits_{j = 2}^{n - 1}
{\left| {i\,j} \right\rangle \left\langle {i\,j} \right|} + \beta
\left( {\left| {00} \right\rangle \left\langle {00} \right| + \left|
{11} \right\rangle
\left\langle {11} \right|} \right) \nonumber\\
&&+ \frac{\beta + \gamma }{2}\left( {\left| {01} \right\rangle
\left\langle {01} \right| + \left| {10} \right\rangle \left\langle
{10} \right|} \right) + \frac{\beta - \gamma }{2}\left( {\left| {01}
\right\rangle \left\langle {10} \right| + \left\langle {10}
\right|\left\langle {01} \right|} \right),
 \end{eqnarray}

\noindent is the state
\begin{eqnarray}
 \sigma ^ * &=& \alpha \sum\limits_{i = 0}^1 \sum\limits_{j = 2}^{n - 1} {\left|
{i\,j} \right\rangle \left\langle {i\,j} \right|} \; + \frac{3\beta
+ \gamma }{6}\left( {\left| {\phi ^ + } \right\rangle \left\langle
{\phi ^ + } \right| + \left| {\phi ^ - } \right\rangle \left\langle
{\phi ^ - } \right| + \left| {\psi ^ + } \right\rangle \left\langle
{\psi ^ + } \right|} \right)\nonumber
\\
 &&+ \frac{3\beta + \gamma }{2}\left| {\psi ^ - } \right\rangle
\left\langle {\psi ^ - } \right|.
 \end{eqnarray}
\noindent where ${|ij\rangle: i=0,1,j=0,1, \cdots, n-1}$.

Our proof goes as follows: if $\sigma ^
* $ is the closest separable state of entangled state $\rho $, then
the value of differential coefficient $\frac{d}{dx}S\left( {\rho
\left\| {\left( {1 - x} \right)\sigma ^ * + x\sigma } \right.}
\right)$ is non-negative, where $\sigma $ is any separable state.
However, if $\sigma^*$ was not a minimum the above gradient would be
strictly negative which is a contradiction.

Proof.  For any given positive operator $A$, we have
\begin{equation}
\label{eq1} \log A = \int_0^\infty {\frac{At - 1}{A +
t}\,\frac{dt}{1 + t^2}}.
\end{equation}

\noindent Let
\begin{equation}
\label{eq2} f\left( {x,\sigma } \right) = S\left( {\rho \left\|
{\left( {1 - x} \right)\sigma ^ * + x\sigma } \right.} \right),
\end{equation}

\noindent  then
\begin{eqnarray}
 \frac{\partial f}{\partial x}\left( {0,\sigma } \right) &=& - \mathop {\lim
}\limits_{x \to 0} \rm tr\left\{ {\frac{\rho \left( {\log \left(
{\left( {1 - x} \right)\sigma ^ * + x\sigma } \right) - \log \sigma
^ * } \right)}{x}} \right\}\nonumber
\\
 &=& \rm tr\left( {\rho \int_0^\infty {\left( {\sigma ^ *
+ t} \right)^{ - 1}\left( {\sigma ^ * - \sigma } \right)\left(
{\sigma ^
* + t}
\right)^{ - 1}} dt} \right) \nonumber\\
 &=& 1 - \int_0^\infty {\rm tr\left( {\rho \left( {\sigma
^ * + t} \right)^{ - 1}\sigma \left( {\sigma^ * + t} \right)^{ - 1}}
\right)}
dt \nonumber\\
 &=& 1 - \int_0^\infty {\rm tr\left( {\left( {\sigma ^ * +
t} \right)^{ - 1}\rho \left( {\sigma ^ * + t} \right)^{ - 1}\sigma }
\right)} dt
\end{eqnarray}

\noindent By Choosing the appropriate basis sequence, $\sigma ^
* $ and $\rho $ can be expressed as the following matrix form
\begin{equation}
 \sigma ^ * = \left( {{\begin{array}{*{20}c}
 {A_{\,1} }  & 0  \\
 0  & {A_{\,2} }  \\
\end{array} }} \right)
, \rho = \left( {{\begin{array}{*{20}c}
 {B_{\,1} }  & 0  \\
 0  & {B_{\,2} }  \\
\end{array} }} \right).
\end{equation}

\noindent where
\[
A_{\,1} = \left( {{\begin{array}{*{20}c}
 {\displaystyle\frac{3\beta + \gamma }{6}}  & 0  & 0  & 0  \\
 0  & {\displaystyle\frac{3\beta + \gamma }{3}}  & { - \displaystyle\frac{3\beta + \gamma
}{6}}  & 0  \\
 0  & { - \displaystyle\frac{3\beta + \gamma }{6}}  & {\displaystyle\frac{3\beta + \gamma
}{3}}  & 0  \\
 0  & 0  & 0  & {\displaystyle\frac{3\beta + \gamma }{6}} \\
\end{array} }} \right),
\]

\[
B_{\,1} = \left( {{\begin{array}{cccc}
 \beta  & 0  & 0  & 0  \\[6pt]
 0  & {\displaystyle\frac{\beta + \gamma }{2}}  & {\displaystyle\frac{\beta - \gamma }{2}}
 & 0  \\[6pt]
 0 & {\displaystyle\frac{\beta - \gamma }{2}}  & {\displaystyle\frac{\beta + \gamma }{2}}
 & 0  \\[6pt]
 0  & 0  & 0 & \beta  \\[6pt]
\end{array} }} \right),
\]

\begin{equation}
 A_{\,2} = \left( {{\begin{array}{*{20}c}
 \alpha  & \ldots  & 0  \\
 \vdots  & \ddots  & \vdots  \\
 0  & \cdots  & \alpha  \\
\end{array} }} \right)£¬
, B_{\,2} = \left( {{\begin{array}{*{20}c}
 \alpha  & \ldots  & 0  \\
 \vdots  & \ddots & \vdots  \\
 0  & \cdots  & \alpha  \\
\end{array} }} \right).
\end{equation}

\noindent $ A_{\,2} ,\,B_{\,2}$ is $\left( {2n - 4} \right)\times
\left( {2n - 4} \right)$ diagonal matrices with diagonal element
$\alpha$, then
\begin{eqnarray}
\fl \left( {\sigma ^ * + t} \right)^{ - 1}\rho \left( {\sigma ^ * +
t} \right)^{ - 1} &=& \left( \begin{array}{cc}
 {\left( {A_{\,1} + t} \right)^{ - 1}}  & 0  \\[6pt]
 0  & {\left( {A_{\,2} + t} \right)^{ - 1}}  \\[6pt]
\end{array}  \right).\left( \begin{array}{cc}
 {B_{\,1} }  & 0  \\[6pt]
 0 \hfill & {B_{\,2} } \\[6pt]
\end{array} \right).\left( \begin{array}{cc}
 {\left( {A_{\,1} + t} \right)^{ - 1}}  & 0  \\[6pt]
 0  & {\left( {A_{\,2} + t} \right)^{ - 1}}  \\[6pt]
\end{array}  \right) \nonumber\\
&=& \left( \begin{array}{cc}
 {\left( {A_{\,1} + t} \right)^{ - 1}B_{\,1} \left( {A_{\,1} + t} \right)^{
- 1}}  & 0  \\[6pt]
 0  & {\left( {A_{\,2} + t} \right)^{ - 1}B_{\,2} \left( {A_{\,2} + t}
\right)^{ - 1}}  \\[6pt]
\end{array}  \right) .
\end{eqnarray}

\noindent Let
\begin{equation}
\label{eq7} g = \int_0^\infty {\left( {\sigma ^ * + t} \right)^{ -
1}\rho \left( {\sigma ^
* + t} \right)^{ - 1}} dt,
\end{equation}

\noindent through some calculations, we obtain
\begin{equation}
 g = \left( {{\begin{array}{*{20}c}
 {C_{\,1} }  & 0  \\[6pt]
 0  & E  \\[6pt]
\end{array} }} \right).
\end{equation}

\noindent where
\begin{equation}
 C_{\,1} = \left( \begin{array}{cccc}
 \displaystyle\frac{6\beta }{3\beta + \gamma }  & 0 & 0 & 0\\[6pt]
 0  & 1  & {\displaystyle\frac{6\beta }{3\beta + \gamma } - 1}  & 0 \\[6pt]
 0  & {\displaystyle\frac{6\beta }{3\beta + \gamma } - 1}  & 1  & 0 \\[6pt]
 0  & 0  & 0  & {\displaystyle\frac{6\beta }{3\beta + \gamma }} \\[6pt]
\end{array}\right),
\end{equation}

\noindent where $E$ is the identity matrix of $\left( {2n - 4}
\right)\times \left( {2n - 4} \right)$. Because $2\left( {n - 2}
\right)\alpha + 3\beta + \gamma = 1,\;0 \le \alpha \le 1 / \left(
{2n - 4} \right)$,  and $1 / 2 \le \gamma \le 1$, we have
\begin{equation} 0 \le ~\frac{6\beta }{3\beta + \gamma } ~\le
1.\end{equation}

\noindent Let  $ \sigma= \left| \eta \right\rangle \left\langle \eta
\right| \otimes \left| \xi \right\rangle \left\langle \xi \right| $,
where $\left| \eta \right\rangle = \sum\limits_n {a_{\,n} } \left| n
\right\rangle $ and $\left| \xi \right\rangle = \sum\limits_n
{b_{\,n} } \left| n \right\rangle $ are orthogonal normalization
vectors, then
\begin{eqnarray}
 \frac{\partial f}{\partial x}\left( {0,\sigma } \right) - 1 &=& - {\rm tr}\left(
{g\sigma} \right) \nonumber\\
 &=& - [\frac{6\beta }{3\beta + \gamma }\left(
{\left| {a_{\,0} } \right|^2\left| {b_{\,0} } \right|^2 + \left|
{a_{\,1} }
\right|^2\left| {b_{\,1} } \right|^2} \right)\nonumber \\
 &&+ \left( {\frac{6\beta }{3\beta +
\gamma } - 1} \right)\left( {a_{\,0} a_{\,1}^ * b_{\,1} b_{\,0}^ * +
a_{\,0}
^ * a_{\,1} b_{\,1} ^ * b_{\,0} } \right) \nonumber\\
&&+ \left| {a_{\,0} } \right|^2\left| {b_{\,1} } \right|^2 + \left|
{a_{\,1} } \right|^2\left| {b_{\,0} } \right|^2 + \sum\limits_{i =
0}^1 \sum\limits_{j = 2}^{n - 1} {\left| {a_{\,i} } \right|^2\left|
{b_{\,j} } \right|^2} ],
\end{eqnarray}

\noindent Due to  $\quad 0 \le \frac{6\beta }{3\beta + \gamma } \le
1$, we get
\begin{eqnarray}
%\begin{equation}
%\begin{array}{l}
\fl \left| {\frac{\partial f}{\partial x}\left( {0,\sigma } \right)
- 1} \right| &=& \frac{6\beta }{3\beta + \gamma }\left( {\left|
{a_{\,0} } \right|^2\left| {b_{\,0} } \right|^2 + \left| {a_{\,1} }
\right|^2\left| {b_{\,1} } \right|^2} \right) + \frac{6\beta
}{3\beta + \gamma }\left( {a_{\,0} a_{\,1}^ * b_{\,1} b_{\,0}^ * +
a_{\,0} ^ * a_{\,1} b_{\,1} ^ * b_{\,0} }
\right) \nonumber\\
  &&+ \left| {a_{\,0} b_{\,1} - a_{\,1}
b_{\,0} } \right|^2 + \sum\limits_{i = 0}^1 \sum\limits_{j = 2}^{n -
1}
{\left| {a_{\,i} } \right|^2\left| {b_{\,j} } \right|^2} \nonumber\\
 &\le& \left( {\left| {a_{\,0} } \right|^2\left|
{b_{\,0} } \right|^2 + \left| {a_{\,1} } \right|^2\left| {b_{\,1} }
\right|^2} \right) + \left( {a_{\,0} a_{\,1}^ * b_{\,1} b_{\,0}^ * +
a_{\,0}
^ * a_{\,1} b_{\,1} ^ * b_{\,0} } \right) \nonumber\\
&& + \left| {a_{\,0} b_{\,1} - a_{\,1} b_{\,0} } \right|^2 +
\sum\limits_{i = 0}^1 \sum\limits_{j = 2}^{n - 1}
{\left| {a_{\,i} } \right|^2\left| {b_{\,j} } \right|^2} \nonumber\\
 &=& \left| {a_{\,0} } \right|^2\left| {b_{\,0}
} \right|^2 + \left| {a_{\,1} } \right|^2\left| {b_{\,1} } \right|^2
+ \left| {a_{\,0} } \right|^2\left| {b_{\,1} } \right|^2  + \left|
{a_{\,1} } \right|^2\left| {b_{\,0} } \right|^2+ \sum\limits_{i
= 0}^1 \sum\limits_{j= 2}^{n - 1} {\left| {a_{\,i} } \right|^2\left| {b_{\,j} } \right|^2} \nonumber\\
 &\le& 1,
%\end{array}
%\end{equation}
\end{eqnarray}

\noindent {\it i.e.}\begin{equation}\frac{\partial f}{\partial
x}\left( {0,\sigma } \right) \ge 0.\end{equation}

For any separable state $\sigma $, which can be expressed by $\sigma
= \sum\nolimits_{\,i} {p_{\,i} \left| {\eta ^i\xi ^i} \right\rangle
\left\langle {\eta ^i\xi ^i} \right|} $, then
\begin{equation}
\frac{\partial f}{\partial x}\left( {0,\sigma } \right) =
\sum\nolimits_{\,i} {p_{\,i} \frac{\partial f}{\partial x}} \left(
{0,\left| {\eta ^i\xi ^i} \right\rangle \left\langle {\eta ^i\xi ^i}
\right|} \right) \ge 0.
\end{equation}
this proves that $\sigma ^ *$ is the closest separable state of
$\rho$ for certain.

\section*{Appendix C}
\appendix
\setcounter{section}{3} Here, we use mathematical induction method
to prove the formula of Fidelity of two-parameter class of states in
$2 \otimes n$ quantum system.

\begin{equation}
F_n = 2\left( {n - 2} \right)\alpha + \frac{3\sqrt {\beta (3\beta +
\gamma )} }{\sqrt 6 } + \frac{\sqrt {\gamma(3\beta + \gamma) }
}{\sqrt 2 }.
\end{equation}

\noindent Proof: when $n=3$, it is easy to obtain the fidelity of $2
\otimes 3$ quantum system,
\begin{equation}
F_3 = 2\alpha + \frac{3\sqrt {\beta (3\beta + \gamma )} }{\sqrt 6 }
+ \frac{\sqrt {\gamma(3\beta + \gamma) } }{\sqrt 2 }.
\end{equation}

\noindent when $n=4$, the Fidelity of $2 \otimes 4$ quantum system,
\begin{equation}
F_4 = 4\alpha + \frac{3\sqrt {\beta (3\beta + \gamma )} }{\sqrt 6 }
+ \frac{\sqrt {\gamma(3\beta + \gamma)} }{\sqrt 2 }.
\end{equation}
Obviously, Eq.(C1) is valid for the cases of $n=3,4$.

Now, let's assume Eq.(C1) is valid for $n =k$, {\it i.e.}
\[
F_{k} = 2\left( {k - 2} \right)\alpha + \frac{3\sqrt {\beta (3\beta
+ \gamma )} }{\sqrt 6 } + \frac{\sqrt {\gamma(3\beta + \gamma)}
}{\sqrt 2 }.
\]

\noindent then when $n =k+1$, by the definition of RGME, we know
\begin{equation}
F_{k+1} = {\rm
tr}\sqrt{\rho^{\frac{1}{2}}\sigma^*\rho^{\frac{1}{2}}}
\end{equation}

\noindent where the matrix expressions of $\rho, \sigma^*$ have been
given in (B6), (B7).

\noindent By calculation, we obtain the Fidelity
\begin{equation}
F_{k+1} =  {\rm tr} \left( \begin{array}{cc}
 V  & 0  \\
 0  & W  \\
\end{array}  \right).
\end{equation}

\noindent where matrix $V$ is a $(2(k+1) - 4)\times (2(k+1) - 4)$
matrix with diagonal element $\alpha $. $W$ is a diagonal matrix
which can be expressed as
\begin{equation}
W = \left( {{\begin{array}{cccc}
 {\sqrt {\displaystyle\frac{\beta \left( {3\beta + \gamma } \right)}{6}} }  & 0
 & 0  & 0  \\
 0  & {\sqrt {\displaystyle\frac{\beta \left( {3\beta + \gamma } \right)}{6}} }
 & 0  & 0 \\
 0  & 0  & {\sqrt {\displaystyle\frac{\beta \left( {3\beta + \gamma }
\right)}{6}} }  & 0  \\
 0  & 0  & 0  & {\sqrt {\displaystyle\frac{\gamma(3\beta + \gamma)
}{2}} } \\
\end{array} }} \right),
\end{equation}

\noindent hence
\begin{eqnarray}
 %F_{k+1} = F_{k} + \left( {2(k+1) - 2\left( {k} \right)} \right)\alpha \\
F_{k+1} &=& 2(k - 1)\alpha + 3\sqrt {\frac{\beta \left( {3\beta +
\gamma } \right)}{6}} + \sqrt {\frac{\gamma(3\beta + \gamma )}{2}} +
\left( {2(k+1) -
2 k } \right)\alpha \nonumber\\
& =&2((k+1)- 2)\alpha + 3\sqrt {\frac{\beta \left( {3\beta + \gamma
}\right)}{6}} + \sqrt {\frac{\gamma(3\beta  + \gamma)}{2}} .
\end{eqnarray}

\noindent That is, when $n=k+1$, Eq.(C1) is also valid. Hence the
proof is over.

\section*{Appendix D}
\appendix
\setcounter{section}{4} Here, we prove the formula of Fidelity of
maximally entangled mixed states in $2 \otimes n$ quantum system is

\begin{equation}
F = \lambda _2 + \lambda _3 + \frac{\left( {\lambda _1 + 2\lambda _4
} \right)}{2}\sqrt {\frac{\lambda _4 }{\lambda _1 + \lambda _4 }} +
\lambda _1 \sqrt {\frac{\lambda _1 + 2\lambda _4 }{2\left( {\lambda
_1 + \lambda _4 } \right)}} ,
\end{equation}
which is independent of $n$.

\noindent Proof: when $n=3,4$, it is easy to obtain the Fidelity of
maximally entangled mixed quantum states by straightforward matrix
calculation.
\begin{equation}
F = \lambda _2 + \lambda _3 + \frac{\left( {\lambda _1 + 2\lambda _4
} \right)}{2}\sqrt {\frac{\lambda _4 }{\lambda _1 + \lambda _4 }} +
\lambda _1 \sqrt {\frac{\lambda _1 + 2\lambda _4 }{2\left( {\lambda
_1 + \lambda _4 } \right)}}.
\end{equation}

%\noindent when $n=4$, the Fidelity of $2 \otimes 4$ maximally
%entangled mixed state is
%\begin{equation}
%F = \lambda _2 + \lambda _3 + \frac{\left( {\lambda _1 + 2\lambda _4
%} \right)}{2}\sqrt {\frac{\lambda _4 }{\lambda _1 + \lambda _4 }} +
%\lambda _1 \sqrt {\frac{\lambda _1 + 2\lambda _4 }{2\left( {\lambda
%_1 + \lambda _4 } \right)}} ,
%\end{equation}

%%%\noindent Obviously, Eq.(D1) is valid for the cases of $n=3,4$.

%%%Now, let's assume Eq.(D1) is valid for $n =k$, {\it i.e.}
%%%\begin{equation}
%%%F_k = \lambda _2 + \lambda _3 + \frac{\left( {\lambda _1 + 2\lambda
%%%_4 } \right)}{2}\sqrt {\frac{\lambda _4 }{\lambda _1 + \lambda _4 }}
%%%+ \lambda _1 \sqrt {\frac{\lambda _1 + 2\lambda _4 }{2\left(
%%%{\lambda _1 + \lambda _4 } \right)}}.
%%%\end{equation}

\noindent We must calculate the Fidelity for different integer $n$,
above formula is still valid.

It is known that matrix $\rho^\prime$ has four eigenvalues
$\lambda_1, \lambda_2, \lambda_3, \lambda_4$. According to Eq.(66),
it can be expressed as

\begin{equation}
\rho^\prime= \left (  \begin{array}{ccc} \lambda_4 &  0  & 0 \\
0 &  P_1 & 0 \\
0 &  0   & P_2 \\
\end{array} \right).
\end{equation}

\noindent where
\begin{equation}
P_1= \left (  \begin{array} {ccc}
\displaystyle \frac{\lambda_1}{2} & 0  &  \displaystyle-\frac{\lambda_1}{2}\\
0     &     R       &    0\\
\displaystyle -\frac{\lambda_1}{2} &  0 & \displaystyle \frac{\lambda_1}{2}\\
\end{array} \right),
P_2= \left(  \begin{array} {ccc}
0 & \ldots & 0  \\
\vdots & \ddots & \vdots \\
0  &  \ldots  &  \lambda_2\\
\end{array}  \right).
\end{equation}

\noindent $P_2$ is a diagonal $(n-1)\otimes (n-1)$ matrix whose last
element of diagonal line is $\lambda_2$. Like matrix $P_2$, $R$ is
$(n-2)\otimes(n-2)$ matrix, but the last element of diagonal line is
$\lambda_3$.

While the matrix expression of $\sigma^\prime$ can be written in the
form

\begin{equation}
\sigma^\prime =\left ( \begin{array}{ccc}
 \displaystyle\frac{(\lambda_1+2
\lambda_4)^2}{4(\lambda_1+\lambda_4)}  &  0  &  0 \\
 0  &  Q_1 &  0  \\
 0  &  0  &  Q_2 \\
 \end{array}  \right).
 \end{equation}

\noindent where
\begin{equation}
\fl ~~~Q_1 =   \left( \begin{array}{ccc}
 \displaystyle\frac{\lambda_1 (\lambda_1+2\lambda_4)}{4(\lambda_1+\lambda_4)}  & 0  &
 \displaystyle-\frac{\lambda_1 (\lambda_1+2\lambda_4)}{4(\lambda_1+\lambda_4)}  \\
 0  &  R  & 0 \\
 \displaystyle-\frac{\lambda_1 (\lambda_1+2\lambda_4)}{4(\lambda_1+\lambda_4)}  & 0  &
 \displaystyle\frac{\lambda_1 (\lambda_1+2\lambda_4)}{4(\lambda_1+\lambda_4)}\\
\end{array}  \right),
Q_2= \left( \begin{array}{cc}
\displaystyle\frac{\lambda_1^2}{4(\lambda_1+\lambda_4)} & 0 \\
0    &     P_2^\prime \\
\end{array}  \right).
\end{equation}

\noindent Same to matrix $P_2$, $P_2^\prime$ is just
$(n-2)\otimes(n-2)$ matrix.

Now $\rho^\prime, \sigma^\prime$ are all expressed in diagonal form,
then we can write
%\begin{equation}
%\rho^\frac{1}{2} \sigma \rho^\frac{1}{2} = \left( \begin{array}{ccc}
%\displaystyle \frac{\lambda_4
%(\lambda_1+2\lambda_4)^2}{4(\lambda_1+\lambda_4)} &
%0 & 0 \\
%0 & P_1^{\frac{1}{2}}  Q_1  P_1^{\frac{1}{2}} &  0  \\
%0 &  0     &  P_2^{\frac{1}{2}} Q_2 P_2^{\frac{1}{2}} \\
%\end{array}  \right)
%\end{equation}

\begin{equation}
F={\rm
tr}\sqrt{{\rho^\prime}^{\frac{1}{2}}\sigma^\prime{\rho^\prime}^{\frac{1}{2}}}={\rm
tr} \left( \begin{array}{ccc} \displaystyle
\frac{\lambda_1+2\lambda_4}{2}\sqrt{\frac{\lambda_4}{
\lambda_1+\lambda_4}} &
0 & 0 \\
0 & \sqrt{P_1^{\frac{1}{2}}  Q_1  P_1^{\frac{1}{2}}} &  0  \\
0 &  0     &  \sqrt{P_2^{\frac{1}{2}} Q_2 P_2^{\frac{1}{2}}} \\
\end{array}  \right).
\end{equation}

\noindent The Fidelity is simplified to
\begin{equation}
 F=\frac{(\lambda_1+2\lambda_4)}{2}\sqrt{\frac{\lambda_4}{
\lambda_1+\lambda_4}}+ {\rm tr}\left(\sqrt{P_1^{\frac{1}{2}} Q_1
P_1^{\frac{1}{2}}}\right)+ {\rm tr}\left( \sqrt{P_2^{\frac{1}{2}}
Q_2 P_2^{\frac{1}{2}}}\right).
\end{equation}

\noindent Here, we use the fact that if a matrix has the form
\begin{equation}
\left( \begin{array}{cccc}
x & 0 & 0 & -x\\
0& O & 0 & 0 \\
0 & 0 & y & 0 \\
-x& 0 & 0 & x \\
\end{array} \right).
\end{equation}
%then the square root of this matrix has the form
%\begin{equation}
%\left(  \begin{array}{cccc}
%\sqrt{x} & 0 & 0 & -\sqrt{x}\\
%0 & 0 & 0 & 0 \\
%0 & 0 & \sqrt{y} & 0 \\
%-\sqrt{x}& 0 & 0 & \sqrt{x} \\
%\end{array} \right).
%\end{equation}

\noindent where $O$ denotes block matrix with all elements 0, then
the eigenvalues of the square root of this matrix are $\sqrt{2x},
\sqrt{y}$.

%Naturally, eigenvalues of matrix (D.11) are $2x,y$, then the
%eigenvalues of matrix (D.10)are $2\sqrt{x},\sqrt{y}$.

In order to get the matrix trace, we start to calculate the
eigenvalues for simplification. Note matrix $P_1^{\frac{1}{2}} Q_1
P_1^{\frac{1}{2}}$ can be expressed in the form (D.10). No matter
what the value of $n$ is, matrix $\sqrt{P_1^{\frac{1}{2}} Q_1
P_1^{\frac{1}{2}}}$ just has two eigenvalues, {\it i.e.} they are
$\lambda_3, \lambda _1 \sqrt {\frac{\lambda _1 + 2\lambda _4 }{2(
\lambda _1 + \lambda _4)}}$. The matrix $P_2^{\frac{1}{2}} Q_2
P_2^{\frac{1}{2}}$ is diagonal matrix with last element
$\lambda_2^2$ in the diagonal line.
%Thus the form of
%$\sqrt{{\rho^\prime}^\frac{1}{2} \sigma^\prime
%{\rho^\prime}^\frac{1}{2}}$ can be easily obtained.

%Finally we can obtain the eigenvalues of $\sqrt{P_1^{\frac{1}{2}}
%Q_1 P_1^{\frac{1}{2}}}, \sqrt{P_2^{\frac{1}{2}} Q_2
%P_2^{\frac{1}{2}}}$.

Thus we acquire the Fidelity according to Eq.(D.9)
\begin{eqnarray}
F &=& \lambda _2 + \lambda _3 + \frac{\left(
{\lambda _1 + 2\lambda _4 } \right)}{2}\sqrt {\frac{\lambda _4
}{\lambda _1 + \lambda _4 }} + \lambda _1 \sqrt {\frac{\lambda _1 +
2\lambda _4 }{2\left( {\lambda _1 + \lambda _4 } \right)}}.
\end{eqnarray}

\noindent It is independent of $n$, that is, when $n$ is an
arbitrary integer, Eq.(D1) is also valid, the proof is finished.

%\noindent where $T$ is a diagonal matrix of $(n-2)\otimes(n-2)$, its
%last element of diagonal line is $\lambda_2$, all other elements are
%zero. $S$ can be expressed as

%where $R$ is a $(n-2)\otimes (n-2)$ matrix whose last element of
%diagonal line is $\lambda_3$, other elements are zero.

%\noindent By calculation, we can obtain the Fidelity

%\begin{equation}
%F_{k+1} =  {\rm tr} \left( \begin{array}{cc}
% Y  & 0  \\
% 0  & Z  \\
%\end{array}  \right).
%\end{equation}

%\noindent where matrix $Y$ is a $(2(k+1) - 4)\times (2(k+1) - 4)$
%matrix with all elements zero. $Z$ is a diagonal matrix which can be
%expressed as
%\begin{equation}
%W = \left( {{\begin{array}{cccc}
% \lambda_2  & 0
% & 0  & 0  \\
% 0  & \lambda_3 & 0  & 0 \\
% 0  & 0  & \displaystyle\frac{\lambda_1+2\lambda_4}{2}\sqrt {\displaystyle\frac{\lambda_4}{\lambda_1+ \lambda_4}}  & 0  \\
% 0  & 0  & 0  & \lambda_1{\sqrt {\displaystyle\frac{\lambda_1+2\lambda_4}{2(\lambda_1+\lambda_4)}} } \\
%\end{array} }} \right),
%\end{equation}

%\noindent hence Fidelity is independent of $k$,
%\begin{eqnarray}
%F_{k+1} &=& \lambda _2 + \lambda _3 + \frac{\left( {\lambda _1 +
%2\lambda _4 } \right)}{2}\sqrt {\frac{\lambda _4 }{\lambda _1 +
%\lambda _4 }} + \lambda _1 \sqrt {\frac{\lambda _1 + 2\lambda _4
%}{2\left( {\lambda _1 + \lambda _4 } \right)}}.
%\end{eqnarray}

%\noindent In other words, when $n=k+1$, Eq.(D1) is also valid, the
%proof is finished.

\section*{References}

\begin{harvard}

\item{[1]} A. Einstein, B. Podolsky and N. Rosen 1935 {\it Phys.
Rev.} \textbf{47} 777

\item{[2]}  E. Schr\"{o}dinger, Naturwissenschaften  1935  \textbf{23} 807-812, 823-828, 844-849.

\item{[3]}  C. H. Bennett, D. P. Divincenzo, J. A. Smolin, and W. K. Wootters   1996 {\it Phys. Rev. A} \textbf{54} 3824

\item{[4]}  Wootters W K  1998  {\it Phys. Rev. Lett.} \textbf{80} 2245

\item{[5]}  Horodecki M  2001 {\it Quantum Inf.Comput.} {\bf 1}, 3  %entanglement measure

\item{[6]}  Wootter W K 2001 {\it Quantum Inf.Comput.} {\bf 1}, 27   %Entanglement of formation and concurrence

\item{[7]}  Vidal G, Dur W and Cirac J  2002  {\it Phys. Rev. Lett.} \textbf{89} 027901

\item{[8]} V. Vedral, M. B. Plenio  1998  {\it Phys. Rev. A} \textbf{57}1619
%quant-ph/9707035

\item{[9]} V. Vedral, M. B. Plenio, M. A. Rippin and P. L. Knight  1998   {\it Phys. Rev. Lett.} \textbf{78}  2275

\item{[10]}  J. Eisert  2001   Ph.D.thesis, University of Potsdam

\item{[11]}  G. Vidal and R. F. Werner  2002   {\it Phys. Rev. A} \textbf{65} 032314

\item{[12]}   G. Vidal and R. Tarrach   1999  {\it Phys. Rev. A} \textbf{59} 141

\item{[13]}  M. Steiner   2003    {\it Phys. Rev. A} \textbf{67}  054305

\item{[14]} Tzu-Chieh Wei and Paul M. Goldbart  2003     {\it Phys. Rev. A} \textbf{68} (4)
042307

\item{[15]} Tzu-Chieh Wei, Joseph B. Alterpeter,  Paul M. Goldbart,  William J.
Munro   2004   {\it Phys. Rev. A} \textbf{70} 022322

\item{[16]} Tzu-Chieh Wei and Paul M. Goldbart 2003  ({\it Preprint} quant-ph/0303079)

\item{[17]} Tzu-Chieh Wei, Marie Ericsson, Paul M.Goldbart and William.J.Munro
2004   {\it Quantum Inform. Compu.} \textbf{4}   252

\item{[18]} Tzu-Chieh Wei and Paul M. Goldbart  2003   ({\it Preprint}
quant-ph/0303158)

\item{[19]}  K. Audenaert, M. B. Plenio and J. Eisert   2003
{\it Phys. Rev. Lett.} \textbf{90} 027901

\item{[20]}  P. M. Hayden, M. Horodecki and B. M. Terhal   2001  {\it J. Phys. A: Math. Gen.} 34
6891

\item{[21]}  E. M. Rains   1999   {\it Phys. Rev. A} \textbf{60} 173

\item{[22]}  M. Horodecki  P. Horodecki and R. Horodecki  1999
 {\it Phys. Rev. A} \textbf{60} 1888

\item{[23]}  S. J. Akhtarshenas M. A. Jafarizadeh  2003  ({\it
Preprint} quant-ph/0304019)

\item{[24]}  An Min Wang, ({\it Preprint} quant-ph/0011040)

\item{[25]} An Min Wang,   ({\it Preprint} quant-ph/02030093; {\it
Preprint} quant-ph/0012029)

\item{[26]} A. Shimony, Ann. N. Y.  1995   {\it Acad. Sci} \textbf{755}  675

\item{[27]} H. Barnum and N. Linden  2001  {\it J. Phys. A: Math. Gen.} \textbf{34} 6787

\item{[28]} G. Vidal  2000   {\it J. Mod. Opt} \textbf{47}, 355

\item{[29]} M. Horodecki, P. Horodecki and R. Horodecki   2000   {\it Phys. Rev. Lett.} \textbf{84} 2014

\item{[30]} M. A. Nielsen and I. L. Chuang  2000  Quantum
Computation and Quantum Information  (Cambridge: Cambridge
University Press)

\item{[31]} A. Peres  1993  Quantum theory: {\sl Computation and
Methods\/}  (Kluwer Academic, Dordecht)

\item{[32]}  W. Dur and J. I. Cirac  2001 {\it J. Phys. A: Math. Gen.} {\bf34}  35  6837-6850

\item{[33]}  Scully M O and Zubairy M S 1999 Quantum
Optics (Cambridge: Cambridge University Press)

\item{[34]} Dong Pyo Chif and Soojoon Lee   2003   {\it J. Phys. A: Math. Gen.} \textbf{36}  11503

\item{[35]} Liang Lin-Mei, Chen Ping-Xing, Li Cheng-Zu and Huang
Ming-Qiu    2001   {\it Chin. Phys. Lett} \textbf{18}  3  325

\item{[36]}  Private communication. Si-ping Liu master'thesis and Ling-yun Li bachlor's thesis.

\item{[37]}  Werner R F  1989  {\it Phys. Rev. A} \textbf{67}  052308

\item{[38]}  Chen P, Liang L, Li C and Huang M    2002   {\it Phys. Lett. A} \textbf{295}
175

\item{[39]}  S. Ishizaka and T. Hirshima   2002   {\it Phys. Rev.
A} \textbf{62} 22310

\item{[40]}  Frank Verstraete, Koenraad Audenaert, Tijl De Bie, Bart
De Moor  2001    {\it Phys. Rev. A} \textbf{64} 012316

\item{[41]}  Pawel Horodecki 1997   {\it Phys. Lett. A}  \textbf{232} 333

\item{[42]} Kai Chen, Sergio Albeverio and Shao-Ming
Fei   2005     {\it Phys. Rev. Lett.} \textbf{95} 040504

\item{[43]}  P. Horodecki, M. Horodecki and R. Horodecki  1999   {\it
Phys. Rev. Lett.}  \textbf{82}  1056

\item{[44]}  K. G. H. Vollbrecht and R. F. Werner  2001   {\it Phys. Rev. A} \textbf{64} 062307

\item{[45]}  M. Horodecki, P. Horodecki  1999
{\it Phys. Rev. A} \textbf{59} 4206

\item{[46]}  R A. Bertlmann, K. Durstberger, B. Hiiesmayr and P. Krammer
2005    {\it Phys. Rev. A} \textbf{72} 052331

\item{[47]}  E. M. Rains   1999    {\it Phys. Rev. A} \textbf{60} 179

\item{[48]}  Pranaw Rungta, Carlton M. Caves  2002 ({\it Preprint}
quant-ph/0208002)

\item{[49]}  W. Dur 2001 {\it Phys. Rev. Lett} \textbf{87}  230402

\item{[50]}  Ofer Biham, Michael A. Nielsen and Tobias J. Osborne  2002
{\it Phys. Rev. A} \textbf{65}  062312

\item{[51]} Daniel Shapira, Yishai Shimoni, Ofer Biham  2006
{\it Phys. Rev. A} \textbf{73} 044301

\item{[52]}  Daniel Cavalcanti  2006   {\it Phys. Rev. A} \textbf{73} 044302

\item{[53]} Jennifer L. Dodd and M. A. Nielsen   2002    {\it Phys. Rev.
A} \textbf{66}  044301

\item{[54]}  Daniel Cavalcanti   2006  { \it Phys. Rev. A} \textbf{73} 044302

\end{harvard}

\end{document}